\documentclass[english,11pt,authoryear]{elsarticle} 
\usepackage[T1]{fontenc}
\usepackage{float}
\usepackage{amsmath}
\usepackage{amsthm}
\usepackage{amssymb}
\usepackage{amsfonts}
\usepackage{bbm}
\usepackage{multirow}
\allowdisplaybreaks
\usepackage{graphicx}
\usepackage{natbib}
\usepackage{float}
\usepackage[countmax]{subfloat}
\usepackage{epstopdf}
\usepackage{hyperref}
\hypersetup{colorlinks,allcolors=black}
\usepackage{nomencl}
\makenomenclature
\usepackage{algorithm}
\usepackage{algorithmic}

\usepackage{url}
\usepackage{graphicx}
\graphicspath{{images/}} 
\usepackage{comment}
\usepackage{ragged2e}
\usepackage{pgfplots}
\pgfplotsset{compat=newest}
\usetikzlibrary{plotmarks}
\usetikzlibrary{arrows.meta}
\usepgfplotslibrary{patchplots}
\usepackage{grffile}
\usepackage{amsmath}
\usepackage{autobreak}
\usepackage{appendix} 
\usepackage{appendix}
\usepackage[symbol]{footmisc}

\usepackage{paralist}
\usepackage{tablefootnote}
\usepackage{booktabs}
\usepackage{nomencl}
\usepackage{amsmath}
\usepackage{array}
\usepackage{makecell}
\makenomenclature

\usepackage{epsfig}
\usepackage{subfigure}
\usepackage{amssymb,amsmath,amsfonts,layout,graphicx}
\usepackage{makeidx}
\usepackage{babel}
\usepackage{tikz}
\usepackage{sublabel}
\usepackage{cases}
\usepackage{algorithm}
\usepackage{algorithmic}

\usepackage
[
        letterpaper,
        left=1.91cm,
        right=1.91cm,
        top=1.91cm,
        bottom=2cm,
]
{geometry}

\newtheorem{remark}{Remark}

\DeclareMathOperator*{\argmax}{arg\,max}

\newcommand{\blue}[1]{\textcolor{black}{#1}}

\renewcommand{\nomgroup}[1]{%
  \item[\bfseries
  \ifstrequal{#1}{S}{Set}{%
  \ifstrequal{#1}{F}{Function}{%
  \ifstrequal{#1}{P}{Parameter}{%
  \ifstrequal{#1}{D}{Decision Variable}{}}}}]
}

\begin{document}

\title{\textbf{Synergizing Drone Delivery Order Pooling and Road Network Monitoring through Monitoring-Task Orderization}}

\author[1staddress,2ndaddress]{Yulong Hu}
\ead{yulong.hu@connect.ust.hk}

\author[2ndaddress]{Meng Xu}
\ead{meng.xu@epfl.ch}

\author[1staddress]{Sen Li}
\ead{cesli@ust.hk}

\author[2ndaddress]{Nikolas Geroliminis\corref{mycorrespondingauthor}}
\cortext[mycorrespondingauthor]{Corresponding author}
\ead{nikolas.geroliminis@epfl.ch}

\address[1staddress]{Department of Civil and Environmental Engineering, 
The Hong Kong University of Science and Technology, Hong Kong, China}

\address[2ndaddress]{School of Architecture, Civil and Environmental Engineering (ENAC), 
École Polytechnique Fédérale de Lausanne (EPFL), Lausanne, Switzerland}

\begin{abstract}
This paper investigates the real-time dispatch of a shared drone fleet to jointly perform on-demand food delivery and urban road network monitoring. We consider a courier-drone collaborative setting in which couriers transport orders to designated launchpads and drones complete the final delivery leg to kiosks. Focusing on drone-side operations, we allow each drone to consolidate multiple orders with distinct origin-destination pairs within a single flight while also making monitoring-aware route adjustments to collect real-time traffic information, subject to delivery-time constraints. This creates a challenging joint decision problem that couples dynamic order-to-drone matching, multi-order pooling, route selection, and time-varying road-network monitoring. Difficulty is further amplified by fleet-level competition for shared delivery orders, interdependent monitoring values induced by overlapping trajectories, and uncertainty in future demand and traffic conditions. To address these challenges, we propose monitoring-task orderization, which periodically converts road-network nodes with high congestion and stale information into virtual monitoring orders. Pooling these virtual tasks with food-delivery orders forms a unified heterogeneous task set and transforms the coupled matching-and-routing problem into a single order-level decision process. Building on this abstraction, we formulate a decentralized graph-interdependent Multi-Agent Markov Decision Process and develop Graph Multi-Agent Q-Learning (Graph-MAQL), which captures localized inter-agent dependencies through bipartite match coordination graphs. The resulting agent-task value estimates are used as edge weights in a dynamic heterogeneous bipartite matching program that guarantees globally feasible execution. Experiments using real-world data reveal a strong operational synergy between delivery and monitoring. Monitoring-task orderization improves monitoring performance by 25.1\% with less than a 1\% reduction in delivery performance, while Graph-MAQL improves the aggregate objective by up to 20.8\%, reduces deadline violations by over 40\%, and transfers zero-shot to higher demand intensity without retraining.
\end{abstract}


\maketitle

\section{Introduction}\label{Sec_Intro}
Recent advancements in unmanned aerial vehicle (UAV) technologies, including sophisticated autonomous navigation systems, high-capacity battery packs, and advanced collision-avoidance mechanisms, have significantly enhanced UAVs' operational capabilities, allowing them to achieve extended ranges, longer flight durations, and high cruising speeds without frequent recharging. Such technological progress has been instrumental in driving the growth of the Low-Altitude Economy (LAE)~\citep{hu2025real}, underscoring the potential of drones to revolutionize and improve the sustainability of urban systems.

Among the diverse urban applications of drone technology, drones are increasingly being recognized as transformative tools for both on-demand logistics and infrastructure monitoring. For on-demand food delivery platforms, traditional courier-based delivery models often struggle with traffic congestion, high operational costs, and strict delivery time requirements. To mitigate these challenges, hybrid courier-drone collaborative delivery systems have been proposed and deployed in cities of China by companies like Meituan~\citep{hu2025real, MeituanNews}. In this model, human couriers handle the first leg of deliveries by picking up orders from restaurants and transporting them to strategically located drone launchpads. Drones then complete the last-leg delivery, leveraging their ability to bypass traffic and efficiently reach customers' nearest kiosks via direct flight paths. This courier-drone synergy not only enhances delivery speed and flexibility but also reduces the workload of human couriers and operational inefficiencies caused by urban traffic constraints and congestion.

Parallel to the developments of UAV for urban on-demand logistics, urban road monitoring with drone technologies has also become a critical need in modern cities to enhance road safety, detect infrastructure anomalies, analyze traffic congestion, extract vehicle trajectories and facilitate effective transportation planning~\citep{barmpounakis2020new,kim2023visual,fonod2025advanced}. Traditional road monitoring methods, such as fixed cameras and manual inspections, are often limited in their coverage and adaptability. A large swarm of drones~\citep{maljkovic2025learning} , however, equipped with advanced imaging collection cameras, could offer a dynamic and scalable solution for real-time road network monitoring. Equipped with high-resolution aerial imaging capabilities, these drones enable the application of advanced computer vision technologies to analyze traffic conditions and identify bottlenecks, thereby significantly enhancing the accuracy, efficiency, and timeliness of road network monitoring efforts.

While drone-based food delivery and road monitoring each hold significant potential, their joint implementation remains relatively underexplored~\citep{xu2026knowledge}, creating substantial opportunities for innovation. With recent advancements in drone technology, such as increased payload capacity, faster flying speeds, and extended battery life, drones now possess the capability to simultaneously serve multiple food orders with different origins and destinations through order bundling and assignments, further expanding the possibilities for integrating food delivery and road network monitoring. Crucially, order bundling and road monitoring are not merely compatible but mutually reinforcing. A drone serving a single point-to-point order traces one fixed path and monitors only the segments along it; once orders are bundled, the drone's trip is stretched across multiple pickup and drop-off stops spanning different zones, so each additional order pooled onto the trip incidentally exposes new road segments to the onboard camera at negligible marginal cost, as depicted in the drone fleet block of Figure~\ref{fig_frameworkoverview}. The reverse coupling holds as well: a minor detour taken to revisit a congested or long-unmonitored segment often repositions the drone near a launchpad with a pending order, so that the same detour that refreshes monitoring data can also be absorbed into the next insertion of the bundling process rather than being purely a delivery cost. This two-way reinforcement means the platform need not treat monitoring and delivery as competing for separate trips; instead, monitoring coverage grows organically with the pooling intensity of the delivery workload, while pooling opportunities are, in turn, discovered along monitoring-motivated detours. This dual-purpose framework not only reduces operational costs but also maximizes resource utilization, dynamically balancing tasks based on real-time demand and road network conditions, thereby contributing to smarter, more sustainable urban systems.

However, seamlessly integrating drone-based food delivery pooling with road network monitoring presents several critical operational challenges. Dispatch algorithms must jointly manage order assignment, route planning, task prioritization, and the coordination of hundreds of drones to achieve system-wide delivery and monitoring objectives in dynamic and uncertain urban environments. A central challenge is balancing the interdependent yet competing objectives of the two tasks. Effective monitoring may require drones to make detours to cover critical road segments, while food orders must still satisfy stringent delivery-time requirements. Moreover, deadline violations often result from the cascading effects of earlier decisions, such as inefficient order bundling. Scalability and real-time decision-making add further complexity, as hundreds of drones must be coordinated to serve delivery requests while selecting among tens of thousands of road segments to maintain adequate monitoring coverage. These challenges are compounded by uncertainty in future order arrivals and road congestion patterns. Effective dispatch strategies must therefore evaluate current assignments and detours not only in terms of their immediate outcomes but also in anticipation of future opportunities, making the integrated problem highly dynamic and computationally demanding. Addressing these challenges requires sophisticated and scalable dispatch algorithms that support efficient system operation while unlocking the synergies between these two complementary applications~\citep{maljkovic2025learning,xu2026knowledge}.

Motivated by these considerations, we develop a unified framework for the real-time dispatch of large-scale drone fleets that jointly perform food delivery order pooling and road network monitoring. At the core of the framework is the novel concept of monitoring-task orderization, which places delivery and monitoring tasks within a common decision space and enables them to be coordinated through a unified task-assignment process. Building on this abstraction, we integrate graph-based multi-agent reinforcement learning with constraint-feasible matching to support scalable and coordinated fleet operations under dynamic conditions. We further evaluate the framework using realistic food delivery and road congestion data from Hong Kong to examine the operational synergy between delivery and monitoring and assess the effectiveness, scalability, and adaptability of the proposed approach. The primary contributions of this work are summarized as follows:

\vspace{-0.3cm}
\begin{itemize}
    \item We propose a real-time fleet-level framework that jointly optimizes on-demand drone delivery order pooling and road network monitoring, enabled by a novel monitoring-task orderization mechanism. Departing from existing studies that treat drones as mono-functional assets carrying a single order per trip, we allow each drone to consolidate multiple orders with distinct origin-destination pairs within one flight while simultaneously sensing the road segments it overflies. The central difficulty of this regime is that the platform must resolve two structurally dissimilar decisions—order-to-drone matching over dynamically arriving orders and hop-by-hop routing over a road network with thousands of edges—whose joint space grows exponentially in the fleet size. To resolve this, we introduce monitoring-task orderization, which periodically ranks road network nodes by a monitoring score that jointly reflects congestion severity and information freshness, and instantiates the top-ranked nodes as virtual monitoring orders. These virtual orders are pooled with real food-delivery orders into a unified heterogeneous order set, collapsing the coupled matching-and-routing decision into a single order-level action, while microscopic flight paths are delegated to a deadline-feasible insertion-based router. Building on this abstraction, we formulate the problem as a decentralized, graph-interdependent Multi-Agent Markov Decision Process, in which each drone is modeled as an order-pooling agent that operates under a shared task-assignment policy, balances the competing delivery and monitoring objectives through a carefully designed reward signal, and anticipates future system dynamics.
    \item We develop Graph-MAQL, a graph multi-agent Q-learning technique that captures the interdependence among agents induced by heterogeneous task assignment. Agents in the proposed system are coupled along two dimensions: they compete for the same pool of food orders at nearby launchpads, while the marginal monitoring value of any decision hinges on the concurrent trajectories of the entire fleet. To encode this coupling, we construct for each agent a localized bipartite match coordination graph, over which a multi-layer graph attention encoder aggregates neighbor information into a coordination-aware embedding; concatenated with the agent's own state embedding and the candidate action feature, it yields the value estimate of every feasible agent-task pair. Because the heterogeneous attributes of delivery and monitoring tasks are folded into a unified state-action representation, a single shared value network evaluates both task types on a common scale, allowing the framework to scale to hundreds of agents without task-specific heads. A dynamic heterogeneous bipartite matching program then converts these decentralized evaluations into a conflict-free joint action, serving as the centralized execution stage of the shared policy. To sustain training at fleet scale, we further adopt a fixed-budget graph sampling strategy that bounds per-epoch message aggregation and the representation fluctuations induced by time-varying neighborhood sizes, while leaving the identities and connectivity of sampled neighbors free to evolve.
    \item \blue{The proposed framework is thoroughly validated through a realistic case study using real-world food delivery and road congestion data from the Kowloon region of Hong Kong. Three findings emerge. First, delivery and monitoring prove mutually reinforcing: under a moderately weighted monitoring objective, both link coverage and bottleneck-road monitoring frequency improve while delivery performance does not deteriorate at all, an effect mediated by order pooling, as the monitoring incentive disperses drones away from over-competed launchpad clusters and thereby exposes each drone to a broader set of poolable orders. Second, reward shaping alone saturates rapidly, since monitoring credit accrues only at the nodes a drone happens to reach; monitoring-task orderization overcomes this ceiling by elevating monitoring into an actively schedulable objective, extending the attainable frontier well beyond what reward weighting can reach at negligible delivery cost. Third, against myopic greedy and independent-learning MARL baselines, Graph-MAQL improves both the delivery and the aggregate objectives while substantially reducing deadline violations, and these gains are sustained zero-shot when demand density rises by a quarter, with no retraining required. These results underscore the potential of the proposed framework to unlock the full capabilities of multi-purpose drone fleets serving both on-demand logistics and urban sensing in large-scale urban environments.}
\end{itemize}

The structure of the remaining paper is as follows: Section~\ref{sec_Literature_Review} reviews relevant studies on UAV-enabled on-demand food delivery, the transition from single-purpose drone operations to integrated delivery and monitoring, and graph-based MARL for mobility dispatch. Section~\ref{Sec_Formulation} describes the problem setting of synergized drone delivery pooling and road network monitoring, presenting first a centralized formulation that exposes its computational intractability, and then our decentralized graph-interdependent MAMDP formulation with monitoring-task orderization and the dynamic heterogeneous bipartite matching process. Section~\ref{sec_GraphMARL} introduces our solution methodology, detailing the construction and encoding of the localized bipartite matching graph and the architecture of the Graph-MAQL framework, while the detailed graph-attention operations and complete training procedure are provided in Appendices~\ref{app_gat_details} and~\ref{app:graph_maql_training}, respectively. Section~\ref{sec_Simulation_Results} presents the simulation setup, experimental results, and discussions. Finally, Section~\ref{sec_conclusions} concludes the paper by summarizing the key findings and outlining future research directions.

\section{Literature Review}\label{sec_Literature_Review}
\subsection{Embracing UAV Technologies in On-Demand Food Delivery}\label{subsec_Drone_Delivery_Review}
Advances in UAV technology—including extended flight ranges, higher cruising speeds, and improved operational reliability—have made drones an increasingly viable option for last-mile logistics, spurring a growing body of research on their application to on-demand food delivery\citep{lyu2025urban}. This literature has evolved along three progressively integrated directions: from drone-only systems, to co-sourcing systems in which drones and human couriers serve as substitutes, and most recently to multimodal systems in which the two modes act as complements, jointly completing a single order through intermodal transfers.

The earliest efforts envisioned drones as a standalone delivery mode and focused on optimizing their operations. \cite{liu2019optimization} cast the fleet management problem as a Mixed Integer Linear Program (MILP) tailored to practical business requirements and designed an algorithm for real-time dispatching. Taking an infrastructure perspective, \cite{pinto2020network} examined how the siting of charging stations shapes system performance and proposed an optimization framework that trades off deployment costs against demand coverage. \cite{levin2023branch} focused on delivery planning within urban airspace, formulating an Integer Linear Program (ILP) and developing a branch-and-price algorithm capable of handling large-scale instances. Recognizing that on-demand food delivery is inherently uncertain, \cite{chen2021improved} introduced a Markov Decision Process (MDP) to address the stochastic optimization problem of planning drone delivery services for fulfilling online and stochastic demand. They developed a supervised learning approach to predict relative link priorities based on the network's state, which were then integrated into an ILP formulation to solve the stochastic optimization problem. The neural network was trained using look-ahead simulation techniques, enabling effective handling of dynamic and uncertain delivery environments.

However, standalone drone operations remain constrained by restricted airspace and the difficulty of serving dense urban environments. These limitations have motivated a second line of work on co-sourcing, which integrates drones into existing courier-based delivery systems: each order is assigned to either a drone or a human courier according to their comparative advantages, with drones favored for long-distance or time-sensitive deliveries where their speed is most valuable, and couriers offering the flexibility to pool multiple orders and to serve areas where drone operations are restricted\citep{chen2026truck}. Methodological contributions in this stream have developed along two distinct directions. From a stochastic control standpoint, \cite{pei2023drone} proposed three models for co-sourcing operations—pure stability, pure profit, and a hybrid variant—that jointly determine the waiting space threshold and the drone fleet size. Leveraging fluid approximations and optimal control theory, they characterized optimal decisions as functions of the outsourcing probability and time delay. From an operations research standpoint, \cite{hamid2023mixed} formulated co-sourcing as an extension of the mixed closed–open pickup-and-delivery vehicle routing problem and solved it with a multi-objective hyper-heuristic that integrates genetic algorithms with a modified particle swarm optimization procedure.

\blue{With recent innovations in food-delivery operations, such as Meituan's pioneering hybrid courier-drone delivery model~\citep{MeituanNews}, a nascent stream of research has begun to investigate the coordinated use of drones and human couriers in multimodal delivery networks with intermodal transfers. This setting differs from the dominant paradigms in the drone-delivery literature, which have largely considered either stand-alone drone operations or co-sourcing systems in which drones and human couriers serve as alternative fulfillment modes and independently complete individual deliveries. In transfer-based courier-drone systems, by contrast, the two modes play complementary roles within the same delivery chain, allowing drones to be deployed without requiring them to independently overcome the first- and last-mile limitations of end-to-end delivery. This operational structure introduces coordination across multiple delivery stages, transportation modes, and resource types: order consolidation and courier operations affect subsequent transfers and drone deployment, while downstream resource availability in turn shapes upstream decisions. Such interactions must typically be managed dynamically under uncertain order arrivals and system states while maintaining stringent delivery-time requirements. These features make the problem structurally different from conventional drone-only and substitute-based co-sourcing settings and have motivated a growing literature on integrated, real-time courier-drone coordination.}

Among the first efforts to close this gap, \cite{hu2025real} explicitly model real-time coordination between couriers and drones in a multimodal, transfer-based delivery pipeline through a Multi-Stage Risk-Aware Multi-Agent Reinforcement Learning (MSRA-MARL) framework. It formulates courier and drone operations as a MAMDP with decentralized Independent RL policies and uses dynamic risk-aware bipartite matching for order assignment. Its Graph Coordinator constructs a spatiotemporal launchpad graph to predict deadline-violation risks caused by transfer delays, incorporating these estimates into both training rewards and matching constraints. To avoid mutually reinforcing suboptimal policies, the authors adopt a curriculum-style procedure that pretrains courier and drone policies under simplified settings before jointly fine-tuning all components with the risk estimator. Complementing the reinforcement-learning-based approach, \cite{dong2026real} tackled the problem from an optimization-based perspective, formulating the joint problem of order-to-courier matching, courier routing and scheduling, and drone repositioning as a MILP under a rolling-horizon re-optimization policy. To address the resulting computational intractability, they proposed an iterative solution algorithm that decomposes the problem into a hypergraph representing feasible courier-task bundle matches and a time-space network tracking drone availability, which are jointly resolved via a column-and-row generation procedure that incrementally identifies feasible many-to-many courier-order matches. \cite{sun2026integrated} formulated the spatiotemporally coupled assignment-and-routing problem as a mixed-integer linear program and embeded it within a rolling-horizon framework for real-time execution, addressing the resulting computational burden with an improved adaptive large neighborhood search (ALNS) algorithm featuring customized removal and insertion operators tailored to the joint courier-drone route structure. 

While the above works have convincingly demonstrated the operational value of UAVs in on-demand food delivery, they share two restrictive premises. First, each drone is assumed to carry a single order per trip, leaving unexplored the consolidation of multiple orders with distinct origin-destination pairs onto one flight, even though advances in payload capacity and multi-compartment designs increasingly render such drone-level pooling technically feasible. Second, drones are cast as mono-functional assets dedicated exclusively to on-demand food delivery: as they traverse the urban airspace, they naturally overfly road segments and could collect real-time traffic information at negligible marginal cost with cameras, yet this sensing capability is entirely overlooked. Relaxing these two premises promises markedly higher fleet utilization and richer service value, but induces a substantially more intricate decision problem in which multi-order routing, cross-task assignment, and heterogeneous operational constraints/objectives must be coordinated jointly in real time. Existing frameworks, architected around single-order, single-task operations, are structurally ill-suited to this regime, leaving a critical gap that our work seeks to fill.

\subsection{From Single-Purpose Drone Operations to Integrated Delivery and Monitoring}\label{subset_UAV}
Beyond their role in logistics, drones have been increasingly adopted for a broad range of urban services that exploit their mobility and sensing capabilities~\citep{mohamed2020unmanned,maljkovic2025learning}. Despite their apparent diversity, these services can be broadly characterized by two fundamental functions: the delivery of physical goods and the acquisition of timely information through onboard sensing. Delivery-oriented services, including parcel, food, and medical delivery, use drones as mobile carriers, whereas sensing-oriented services leverage them as flexible information-collection platforms: infrastructure inspection and public-safety surveillance are, at their core, monitoring tasks in which drones repeatedly observe assets or areas to detect anomalies and track evolving conditions \citep{fang2024strategies}, while aerial surveying and mapping use onboard sensing to acquire spatial information about the environment \citep{colomina2014unmanned}. Reflecting this functional distinction, existing research has primarily developed along two parallel tracks: drone-enabled delivery and drone-based monitoring, each treating the drone as a mono-functional asset dedicated to a single purpose. This separation, however, is largely an artifact of divergent research trajectories rather than an inherent technological limitation, since a single fleet can in principle serve both functions concurrently.

A growing body of work has begun to bridge these two functions by tasking delivery drones with monitoring duties, along a spectrum of increasing coupling. At one end, monitoring is incorporated as an auxiliary capability of otherwise delivery-driven operations without reshaping the underlying delivery route. \cite{xiang2023reusing} reused delivery drones for urban monitoring by selecting, among candidate delivery paths, those most favorable for monitoring while jointly tuning monitoring locations and payload allocation. Similarly, \cite{xiang2023enabling} exploited delivery drones for wireless data collection by optimizing package-route selection together with data-collection time allocation. In both cases, monitoring benefits are extracted from delivery-compatible trajectories, while the routing space itself remains determined by the logistics task. Along a different axis, \cite{liu2025delay} kept the delivery route fixed but introduced online execution control: for a drone on a predetermined mission, a finite-horizon MDP decides in real time how fast to fly and when to pause for in-situ sensing, letting monitoring influence how the mission is carried out in time without altering its route. A tighter integration retains delivery as a primary mission but allows shared drones to be dispatched or locally rerouted to accomplish monitoring tasks. For example, \cite{gao2024sharing} determined which instant-delivery drones should be shared for monitoring fixed points of interest and responding to stochastic anomaly events, together with when and how long they should monitor, while preserving delivery priorities. 

With tighter integration, monitoring is elevated from an auxiliary capability to an explicit routing objective. \cite{liu2023optimal} formulated the joint goods delivery and in-situ monitoring problem as a finite-horizon Markov decision process that jointly selects delivery actions and monitoring waypoints for a single drone under delivery deadlines. \cite{liu2023decentralized} extended this formulation to multiple drones through decentralized game-theoretic routing that balances monitoring rewards, delivery deadlines, and trajectory privacy. \cite{campbell2023multipurpose} cast joint delivery and monitoring as a multi-drone general routing problem, jointly routing a fleet to serve delivery locations while providing coverage over continuous sensing regions so as to minimize total travel duration. More recently, \cite{lyu2026multi} developed a multi-service network design framework in which multi-function drones jointly provide urban monitoring, data collection, and one-to-one parcel pickup-and-delivery services, with drone routes and data-transmission decisions optimized simultaneously via mixed integer quadratically constrained formulations. \cite{xu2026knowledge} let monitoring value justify detours, allowing each drone to deviate from its delivery path to overfly road segments of interest under a knowledge-aware Monte Carlo Tree Search that balances coverage gains against detour cost. Within these studies, routing becomes a genuinely joint delivery-monitoring decision, representing an important step towards multi-purpose drone operations. 

Existing studies have convincingly demonstrated that delivery and monitoring can be jointly supported by a shared drone fleet. The remaining challenge is therefore no longer whether these two services can coexist, but how they should be coordinated when delivery itself is modeled as a realistic on-demand operation. In practical on-demand delivery systems, delivery involves a sequence of tightly coupled operational decisions, including order-to-drone matching, the pooling of multiple heterogeneous orders onto a single flight, the routing complexity induced by such consolidation, and the coordination between conflicting objectives of on-demand drone delivery pooling and monitoring task satisfaction. Addressing this challenge requires a real-time, fleet-level operational framework that jointly coordinates these delivery decisions with the allocation of drone capacity between delivery and monitoring under stochastic demand.

\subsection{Graph MARL and its Application in Mobility Dispatch} 
Graph Neural Networks (GNNs) have emerged as a powerful class of models for learning over relational data, with representative architectures such as the Graph Convolutional Network (GCN)~\citep{kipf2016semi}, GraphSAGE~\citep{hamilton2017inductive}, and the Graph Attention Network (GAT)~\citep{velivckovic2017graph} demonstrating remarkable computational efficiency and representational power in complex and adaptive representation learning tasks. As multi-agent systems are naturally graph structured with agents as nodes and their interactions as edges, researchers have increasingly sought to integrate these potent GNN models into MARL frameworks. This nascent line of research is motivated by several challenges inherent to MARL that GNNs are well positioned to address: encoding environmental dynamics from the perspective of individual agents whose observations are partial and relational, capturing the time-varying interaction topology among agents, supporting the decomposition of joint value functions along with the nuanced assignment of credit across the collective team, and—critically—achieving scalability to large agent populations, as the permutation invariance and parameter sharing intrinsic to GNNs allow policies to generalize across varying numbers of agents while localized message passing circumvents the combinatorial explosion of the joint state–action space~\citep{munikoti2023challenges,11125816}.

Within the specific context of multi-agent coordination games, existing work of graph MARL has advanced along two complementary directions: learning abstract interaction representations among agents, and structuring value factorization through explicit graph topologies. Along the first direction, \cite{jiang2018graph} proposed graph convolutional reinforcement learning (DGN), which models the multi-agent environment as a dynamic graph and employs multi-head attention as the convolution kernel to extract relational features from each agent's local neighborhood. By stacking convolutional layers, the receptive field of each agent progressively expands, allowing cooperation to emerge over increasingly larger scopes, while a temporal relation regularization term encourages consistency of attention weights across consecutive time steps to stabilize cooperative behavior. Building on this idea, \cite{liu2020multi} introduced a two-stage attention mechanism (G2ANet) for game abstraction: a hard-attention stage first prunes irrelevant agent pairs to determine whether an interaction exists, after which a soft-attention stage assigns fine-grained importance weights to the remaining interactions. This two-stage design substantially reduces the complexity of learning in large-scale agent populations and can be flexibly combined with both policy-based and value-based MARL algorithms. Moving beyond attention-based weighting over a predefined topology, \cite{duan2026bayesian} treated the interaction structure itself as a latent variable and proposed a Bayesian ego-graph inference framework for networked MARL, in which each agent infers a posterior distribution over its local ego-graph and performs message passing over the sampled subgraphs, thereby capturing the uncertainty of inter-agent dependencies and adapting the communication structure to the evolving environment. Along the second direction, \cite{wei2022vgn} leveraged the notion of a coordination graph and utilized GATs to factorize the joint team value function or team policy over the graph structure, so that each agent's contribution is contextualized by its neighbors and coordinated behaviors among agents can be explicitly induced. \cite{duan2024inferring} inferred a latent temporal sparse coordination graph directly from agents' observation trajectories, simultaneously capturing the uncertainty and temporal dynamics of inter-agent dependencies for message passing over the learned structure.

Encouraged by these advances, several pioneering studies have begun to extend graph-based MARL to real-world mobility dispatching systems. \cite{zhou2022gmix} proposed GMIX, a graph-based spatial–temporal MARL framework for dynamic electric vehicle dispatching, which employs GNNs to encode spatial dependencies among dispatching regions and couples the learned spatiotemporal representations with a QMIX-style value-mixing network, enabling coordinated fleet-level decisions under time-varying charging and demand dynamics. Focusing on ride-pooling order dispatch, \cite{11125816} developed BMG-Q, a localized bipartite match graph attention Q-learning framework that dynamically constructs a bipartite graph between vehicles and candidate orders at each decision epoch, where a GAT-enhanced deep Q-network captures local interdependencies among neighboring agents to refine vehicle–order value estimates before combinatorial matching, achieving scalability to thousands of agents while mitigating the overestimation bias of independent Q-learning. More recently, \cite{tresca2026robo} demonstrated robo-taxi fleet coordination at scale, leveraging graph-network-based policies defined over the city topology to jointly manage dispatching and rebalancing, and showing that such relational inductive biases yield robustness and transferability across cities and fleet sizes.

Despite these promising developments, existing graph MARL frameworks for mobility dispatch are designed for a single, homogeneous task class: every agent draws its actions from one type of task, so that the coordination graph, the value representation, and the downstream assignment program all operate under a single task semantics. None of them accommodates a fleet whose agents must arbitrate, at every decision epoch, among structurally dissimilar tasks that carry distinct attributes, feasibility constraints, and reward mechanisms, yet compete for the same finite fleet capacity. Following this line of research, we extend graph-based MARL to a heterogeneous task space, in which delivery and monitoring tasks are folded into a unified state-action representation so that a single shared value network prices both task types on a common scale, while the coordination graph continues to encode the contention among agents induced by their overlapping candidate task sets. A dynamic heterogeneous bipartite matching program then converts these value estimates into a conflict-free joint action, enabling coordinated decision-making across dissimilar tasks within a unified architecture.

\section{Problem Description and Formulation}\label{Sec_Formulation}
To synergize drone delivery with road network monitoring and study their joint potential, we consider a hybrid courier-drone delivery mode facilitated by enabling infrastructures such as launchpads and kiosks. In this system, drones collaborate with human couriers to complete food delivery orders within a multimodal delivery network. Human couriers manage the first leg of the delivery by picking up food orders from restaurants. After pickup, if the remaining distance of the order is appropriate, the courier is dispatched to transfer the order to the nearest launchpad, which serves as an intermediate hub for order handovers. Launchpads function as airports for drones, where food orders are loaded onto drones for the subsequent flight leg. Throughout this work, the courier leg is treated as exogenous: orders whose remaining distance is suitable for drone service are assumed to reach the nearest launchpad without delay, so that our decision scope begins at the launchpad. This abstraction isolates the drone-side dispatch problem, which is the main focus of this paper, while real-time courier-drone coordination under transfer uncertainty has been addressed separately in other works ~\citep{hu2025real}.

\begin{figure}
    \centering
    \includegraphics[width=0.75\linewidth]{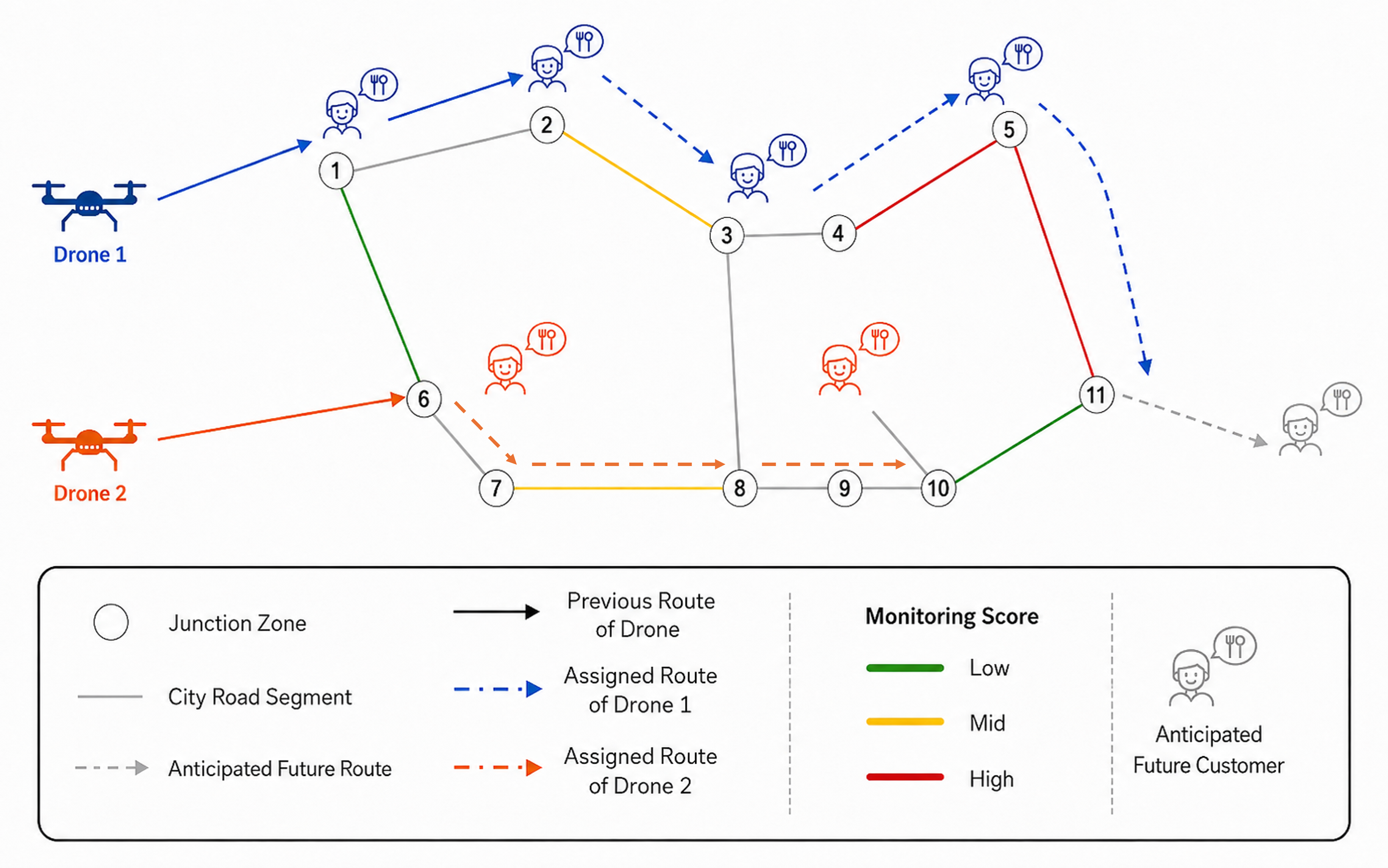}
    \caption{\blue{An illustrative example of dispatching drone fleets for on-demand delivery pooling and road network monitoring.}}
    \label{fig_toyexample}
\end{figure}

\blue{Once loaded, drones depart from the launchpads and deliver the orders to the kiosks closest to the customers. A kiosk is an automated device that stores multiple food orders and enables customers to retrieve their orders by scanning a barcode. Different from the drone settings introduced in~\cite{hu2025real}, our setting extends drone operations in two important directions. First, each drone can serve multiple orders with distinct origin-destination pairs within a single trip, in a manner analogous to ride-pooling: beyond loading several orders at its departure launchpad, a drone already en route may be dynamically assigned new orders and stop at intermediate launchpads to pick them up before completing its remaining deliveries. Second, drones are no longer dedicated exclusively to delivery. As they traverse the urban airspace, they may deliberately route toward road-network nodes of interest and collect real-time traffic information from the road segments incident to the visited nodes, allowing each flight to jointly serve delivery and road-network monitoring tasks. An illustrative toy example is provided in Figure~\ref{fig_toyexample}. Upon observing incoming customer requests and the current congestion status of the road network, the platform jointly determines delivery assignments and monitoring-aware routes for the drone fleet. As illustrated, Drone 1 and Drone 2 are dispatched to different regions of the network and may sequentially serve multiple customers while traversing a set of junction zones. Importantly, the selected routes need not coincide with the shortest delivery paths: since visiting a junction enables the drone to refresh traffic information on its incident road segments, the platform may intentionally route a drone through junctions associated with higher monitoring scores, provided that the resulting detour does not excessively compromise delivery timeliness. At the same time, anticipated future customer requests influence downstream fleet positioning, creating incentives to move drones toward locations that are promising for subsequent order fulfillment. Consequently, each routing decision must balance current delivery service, the freshness and spatial coverage of traffic information, and the positioning of the fleet for future demand, making the resulting problem inherently anticipatory and multi-objective rather than a conventional delivery-only routing problem.}

\blue{To realize this synergy, the platform must orchestrate the drone fleet through a series of intricate real-time decisions, including order bundling, task assignment, route selection, and drone repositioning, all made in the face of substantial uncertainty arising from future order arrivals, evolving road congestion patterns, and the decaying freshness of collected traffic information. The platform's objective is to maximize delivery profit through timely order completion, while simultaneously improving road-network coverage and maintaining frequent monitoring of critical road segments. A key challenge is that delivery and monitoring tasks are intrinsically heterogeneous: delivery requests are explicit origin-destination jobs with service deadlines, whereas monitoring demand is spatially distributed over the road network and evolves continuously with traffic conditions and information freshness. The challenge is further amplified at scale, as dense road networks and large drone fleets induce a combinatorial explosion in feasible delivery–monitoring assignments and routing choices, making direct joint optimization increasingly difficult in real time. To bridge this structural mismatch, we introduce a \textbf{monitoring task orderization} technique that converts dynamically emerging monitoring needs into order-like tasks that can be represented, evaluated, and assigned together with delivery requests within a unified dispatching framework. This transformation enables the platform to explicitly trade off the delivery value and monitoring value of alternative drone movements, rather than treating traffic sensing merely as an incidental by-product of delivery routes. The remainder of this section first presents a centralized mathematical formulation of the problem to expose its computational intractability in subsection~\ref{sec_idealformulation}, and then develops our decentralized graph-interdependent MAMDP formulation, in which the proposed monitoring task orderization technique and heterogeneous bipartite matching graphs jointly support scalable real-time coordination, in subsection~\ref{sec_realformulation}.}

\subsection{Centralized Formulation}\label{sec_idealformulation}
We divide the city into several traffic analysis zones, represented by the set \( \mathcal{B} \). At each decision time \( t \in \mathcal{T} = \{0, \Delta t, 2\Delta t, \dots, T\} \), where \( \Delta t \) denotes the decision interval and \( T \) is the finite planning horizon, the system state captures the real-time status of all orders, couriers, and drones in the city. In particular, we denote the system state as \( S_t \), which integrates the spatial and operational information of unmatched orders, orders awaiting drone delivery, couriers, and drones.

The set of unmatched food delivery orders at time \( t \), denoted by \( \mathcal{O}_{u,t} \), includes all orders that have been placed but have not yet been assigned to a courier. Each order \( o \in \mathcal{O}_{u,t} \) is characterized by:
\begin{equation}
\label{eq:new_order_info}
    p_{o_u} = (b_{\text{origin}}, b_{\text{dest}}, t_s, t_d),
\end{equation}
where \( b_{\text{origin}} \in \mathcal{B} \) is the originating zone of the order, \( b_{\text{dest}} \in \mathcal{B} \) is the destination zone, \( t_s \in \mathbb{R} \) is the time the order was placed, and \( t_d \in \mathbb{R} \) is the expected delivery deadline. Upon observing these orders, those with appropriate delivery distances, constituting a fraction $p_r$ of all orders, are assumed to be immediately picked up by couriers and transported to the nearest launchpad $l\in \mathcal{L}$ within the same zone.

The order \( o_l \in \mathcal{O}_{L,t} \) arrived at the launchpad will then be characterized by:
\begin{equation}\label{eq:new_launchpad_order_info}
    p_{o_l} = (b_l, b_{\text{dest}}, t_l, t_d),
\end{equation}
where \( b_l \in \mathcal{B} \) represents the zone of the launchpad $l\in \mathcal{L}$ to which the order has been transported, and \( t_l \in \mathbb{R} \) is the time the order arrived at the launchpad $l$. The set of orders awaiting drone delivery at launchpads at time \( t \), denoted by \( \mathcal{O}_{L,t} \), includes all orders that have been transported to launchpads by couriers but are waiting for drones to complete the final delivery.

The state of each drone \( m \in \mathcal{M} \) at time \( t \), denoted by \( I_{m,t} \), includes its current location and operational status. It is represented as:
\begin{equation}
\label{eq:drone_info}
    I_{m,t} = (b_{m,t}, v_{m,t}, p_{m,t}, t)
\end{equation}
where $b_{m,t}$ encodes the current location; $v_{m,t}$ is the number of vacant capacities; $p_{m,t}$ encodes the information of food orders on board of drone $m$, including their estimated remaining time on board, drop-off locations, and current existing detours; and $t$ is the current time. 

The road network monitoring state at time \( t \), denoted by \( g_{r,t} \), captures the real-time traffic and monitoring status of the urban road network. We model the road network as a directed graph \( \mathcal{G}_r = (\mathcal{N}, \mathcal{E}) \), where \( \mathcal{N} \) is the set of nodes (road intersections) and \( \mathcal{E} \) is the set of edges (road segments). Each edge \( e \in \mathcal{E} \) at time \( t \) is characterized by:
\begin{equation}
\label{eq:edge_info}
    p_{e,t} = \left( v_{e}, v_{\text{limit},e}, t_e \right),
\end{equation}
where \( v_{e} \in \mathbb{R} \) is the real-time traffic speed on edge \( e \) extracted from the fused speed map, \( v_{\text{limit},e} \in \mathbb{R} \) is the speed limit of edge \( e \), and \( t_e \in \mathbb{R} \) is the time elapsed since edge \( e \) was last monitored by any drone. The road network monitoring state is then defined as:
\begin{equation}
\label{eq:road_network_state}
    g_{r,t} = \left( \mathcal{G}_r, \{p_{e,t}\}_{e \in \mathcal{E}} \right),
\end{equation}
which aggregates the topology of the road network together with the real-time congestion levels and monitoring recency of all road segments. As drones traverse the urban airspace and new traffic information is collected, \( g_{r,t} \) is updated synchronously at every decision time, thereby providing the basis for computing the edge-level monitoring scores introduced later in this section.

Combining these components, the global system state at time \( t \) is defined as:
\begin{equation}
    S_t = \left(\{p_{o_l}\}_{o_l \in \mathcal{O}_{L,t}}, \{I_{m,t}\}_{m \in \mathcal{M}}, g_{r,t}, t \right),
\end{equation}
where the system state integrates the information of unmatched orders at launchpads, drones, road network graphs, and the current time step.

Given the system state, the platform must make two major types of control decisions for each drone at every decision time. First, the platform determines order matching decision \( x_{m,o_l,t} \in \{0, 1\} \), whether drone \( m \) is assigned to order \( o_l \in \mathcal{O}_{L,t} \) at time \( t \); in the spirit of ride-pooling, new orders may also be inserted into the ongoing trip of an en-route drone, provided that the drone has vacant capacity and the induced detours remain acceptable. Second, the platform determines routing decisions \( y_{m,n,t} \in \{0, 1\} \), where each drone \( m \) selects the next road network node \( n \in \mathcal{N} \) as its next waypoint at time \( t \), where \( n \) is restricted to the nodes adjacent to the drone's current node \( n_{m,t} \) in the road network graph \( \mathcal{G}_r \). This decision jointly governs both delivery and monitoring performance: the chosen node determines the flight path along which onboard orders are transported toward their destination kiosks, as well as which road edges are traversed and monitored along the way. 

With these decisions being made, the platform's objective is to maximize the expected total profit generated by the drone fleet over the entire planning horizon, jointly accounting for delivery revenue, road network monitoring utility, operational costs, and delivery timeliness:
\begin{equation}
\label{eq:centralized_objective}
    \max_{\{x_{m,o_l,t}\},\ \{y_{m,n,t}\}} \ \mathbb{E} \left[ \sum_{t \in \mathcal{T}} \sum_{m \in \mathcal{M}} \left( B_{m,o_l,t} + \alpha B_{m,z,t} - C_{m,t} - P_{m,t} \right) \right],
\end{equation}
The expectation is taken over the stochasticity of future order arrivals and road traffic evolution. \( B_{m,o_l,t} \) is the delivery bonus earned by drone \( m \) for serving matched orders on time, \( B_{m,z,t} \) is the monitoring bonus earned for collecting real-time traffic information over road edges adjacent to its location, \( \alpha \) is a tunable global monitoring coefficient, \( C_{m,t} \) is the operational cost of utilizing drone \( m \), and \( P_{m,t} \) penalizes detours that jeopardize the delivery deadlines of on-board orders. The four components are specified as follows:
\begin{itemize}
\item \( B_{m,o_l,t} \) is the delivery bonus earned by drone \( m \) for serving newly matched orders at launchpads, calculated as \( B_{m,o_l,t} = \sum_{o_l \in \mathcal{O}_{L,t}} (R_{o_l} + \beta_d \cdot R_{o_l}^d) \cdot x_{m,o_l,t} \), where \( R_{o_l} \) is the platform revenue earned from matching order \( o_l \) to drone \( m \), \( R_{o_l}^d \) is the distance-based bonus, and \( \beta_d \) is the corresponding scaling factor.

\item \( B_{m,z,t} \) is the total monitoring bonus earned by drone \( m \) for the information gain of monitoring all road edges adjacent to its current node, following the spirit of~\citep{maljkovic2025learning}:
\begin{equation}
B_{m,z,t} = \sum_{e \in \mathcal{E}(n_{m,t})} \text{score}(e, t),
\end{equation}
where \( n_{m,t} \) is the current node of drone \( m \) at time \( t \), determined by its routing decisions \( \{y_{m,n,\tau}\}_{\tau \le t} \), and \( \mathcal{E}(n_{m,t}) \) denotes the set of road edges adjacent to node \( n_{m,t} \). For each road edge \( e \), its real-time monitoring score is formulated according the edge attributes $p_{e,t}$ as:
\begin{equation}
\label{eq:monitoring_score}
\text{score}(e, t) =  \left( \alpha_c \cdot \max\left(1 - \frac{v_e}{v_{\text{limit},e}},\ 0.1\right) + (1-\alpha_c) \cdot \left(1 - 0.5^{\frac{t_e}{T_h}}\right) \right),
\end{equation}
where \( v_e \) is the real-time traffic speed on edge \( e \) extracted from the fused speed map, \( v_{\text{limit},e} \) is the speed limit of edge \( e \), \( t_e \) is the time elapsed since edge \( e \) was last visited by \emph{any} drone, and \( \alpha_c \) is the balancing coefficient between congestion severity and information freshness. \( T_h \) controls the characteristic time scale of freshness decay. The first term assigns greater monitoring value to currently congested edges, with a floor value of 0.1 to preserve a baseline incentive for monitoring uncongested roads. The second term captures the declining informativeness of historical observations: an edge monitored recently receives little additional freshness reward, whereas its monitoring value gradually increases as the elapsed time since the last observation grows. \blue{Together, the two terms exploit the temporal persistence of traffic conditions by avoiding redundant repeated observations, while directing the limited sensing capacity of the drone fleet toward road segments that are either operationally critical due to congestion or increasingly uncertain due to stale information.}


\item \( C_{m,t} = C_{m} \cdot w_{m,t} \) is the operational cost incurred by drone \( m \), where \( C_m \) is the per-minute operating cost and \( w_{m,t} \in \{0,1\} \) indicates whether drone \( m \) is utilized at time \( t \).

\item \( P_{m,t} \) is the detour penalty incurred by drone \( m \) for its on-board orders \( o_m \in \mathcal{O}_{m,t} \):
\begin{equation}
\label{eq:detour_penalty}
P_{m,t} = \sum_{o_m \in \mathcal{O}_{m,t}} \lambda_{c} \cdot \min\left(\Delta_{m}^{\text{det}},\ \Gamma_{m}^{\text{det}}\right) + \lambda_{p} \cdot \max\left(\Delta_{m}^{\text{det}} - \Gamma_{m}^{\text{det}},\ 0\right),
\end{equation}
where \( \Delta_{m}^{\text{det}} \) is the extra travel time induced by detours for newly added tasks, \( \Gamma_{m}^{\text{det}} \) is the tolerance threshold for acceptable detours constrained by delivery deadlines, and \( \lambda_c, \lambda_p \) are positive weighting coefficients for mild and excessive detours, respectively. This piecewise penalty enforces on-time delivery and discourages detours that would violate order deadlines.
\end{itemize}

Although Equation~\eqref{eq:centralized_objective} compactly characterizes the platform's goal, solving this centralized problem is computationally intractable for three reasons. First, the joint decision space grows combinatorially with the fleet size and the network scale: the order matching decisions \( \{x_{m,o_l,t}\} \) constitute a set-partitioning problem over dynamically arriving orders across \( |\mathcal{M}| \) drones, coupled multiplicatively with the routing decisions \( \{y_{m,n,t}\} \), whose joint space grows exponentially in the number of drones and is further amplified by the massive number of nodes and edges in a realistic urban road network \( \mathcal{G}_r \). Second, the agents are tightly coupled through two dimensions of competition. On the monitoring side, the freshness term \( t_e \) in Equation~\eqref{eq:monitoring_score} depends on the visitation history of \emph{all} drones, so the marginal monitoring value of any routing decision hinges on the concurrent trajectories of the entire fleet. On the delivery side, drones compete for the shared pool of orders \( \mathcal{O}_{L,t} \) at launchpads: assigning an order to one drone excludes all others, and the profitability of each candidate match depends on the vacant capacities, on-board detour budgets, and locations of competing drones. Third, the objective must be optimized under substantial uncertainty in future order arrivals and traffic evolution, while the reward structure itself embeds competing incentives---pursuing monitoring bonuses \( B_{m,z,t} \) or bundling additional orders \( B_{m,o_l,t} \) through routing detours directly conflicts with the timeliness enforced by \( P_{m,t} \). These difficulties render existing frameworks such as Multi-Agent Path Finding, Vehicle Routing Problem (VRP), and Monte Carlo Tree Search-based planning inadequate, as none simultaneously handles dynamic order arrivals, fleet-level coupling, and the exponential joint decision space. This motivates our decentralized MAMDP formulation empowered by heterogeneous bipartite match assignment graphs and monitoring task orderization, presented next.

\subsection{Decentralized Graph Interdependent MAMDP Augmented with Heterogeneous Bipartite Matching}\label{sec_realformulation}

\begin{figure}
    \centering
    \includegraphics[width=1\linewidth, height = 10cm]{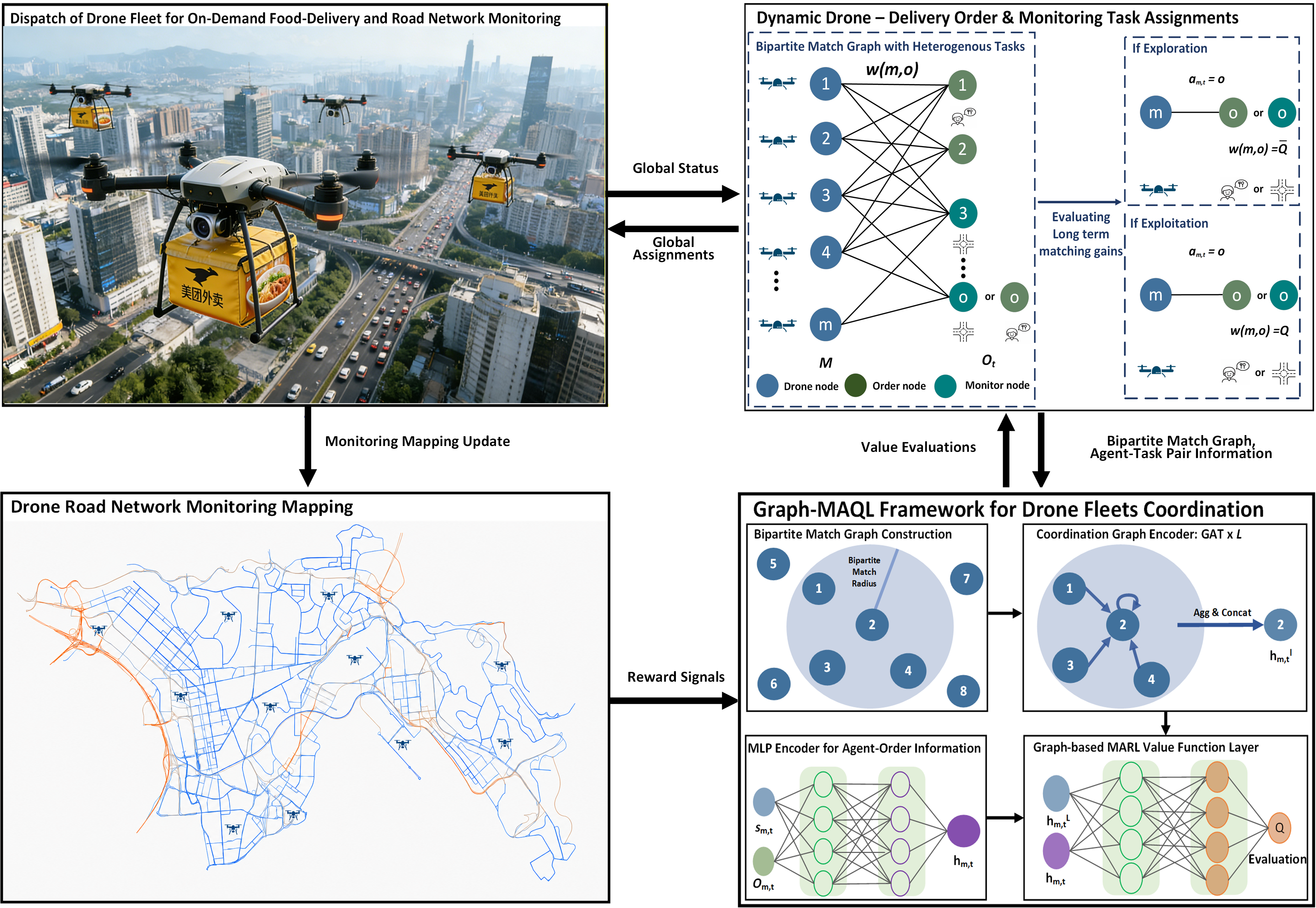}
    \caption{\blue{Overview of our proposed framework for dispatch of drone fleets for on-demand delivery and road network monitoring.}}
    \label{fig_frameworkoverview}
\end{figure}

\subsubsection{Decentralized Graph Interdependent MAMDP Formulation with Monitoring-task Orderization}

In this subsection, we formulate the problem as a Multi-Agent MDP integrated under a dynamic heterogeneous bipartite matching process, where each drone \( m \in \mathcal{M} \) is modeled as an order-pooling agent operating within a grid-based system, as shown in the drone road network monitoring mapping block in Figure~\ref{fig_frameworkoverview}.

The key innovation enabling this formulation is what we term \textbf{\emph{monitoring-task orderization}}, which unifies the two heterogeneous decisions \( x_{m,o_l,t} \) and \( y_{m,n,t} \) in the previous centralized formulation into one single action \( a_{m,t} \). Instead of letting each drone make hop-by-hop routing decisions over the massive road network, we periodically convert the monitoring demand itself into \emph{virtual monitoring orders}. Specifically, every \( T_m\Delta T \) minutes, the platform computes the aggregate monitoring score of every node \( n \in \mathcal{N} \) as defined in Equation~\eqref{eq:monitoring_score}, and sorts the nodes in non-increasing order of this score. Let \( \pi_t : \{1,\dots,|\mathcal{N}|\} \to \mathcal{N} \) denote the resulting ranking permutation at time \( t \), i.e.,
\begin{equation}
\sum_{e \in \mathcal{E}(\pi_t(1))} \text{score}(e,t) \ \geq\ \sum_{e \in \mathcal{E}(\pi_t(2))} \text{score}(e,t) \ \geq\ \cdots \ \geq\ \sum_{e \in \mathcal{E}(\pi_t(|\mathcal{N}|))} \text{score}(e,t).
\label{eq:ranking}
\end{equation}
The platform then selects the top-\(k\) nodes as
\begin{equation}
\mathcal{N}_{Z,t} = \{\pi_t(1), \pi_t(2), \dots, \pi_t(k)\}, \qquad |\mathcal{N}_{Z,t}| = k \ll |\mathcal{N}|.
\label{eq:topk_selection}
\end{equation}
Each selected node \( n \in \mathcal{N}_{Z,t} \) is then instantiated as a virtual monitoring order \( o_z \in \mathcal{O}_{Z,t} \) via the mapping
\begin{equation}
o_z = \big(b_{\text{origin}} = b(n),\ b_{\text{dest}} = b(n),\ t^{o}_s = t,\ t^{o}_d = t + T_m\Delta T,\ \kappa_o = \text{monitor}\big),
\label{eq:virtual_order_map}
\end{equation}
where \( b(n) \) denotes the zone containing node \( n \). Unlike a food order, a virtual monitoring order has no separate pickup and drop-off location since its sole purpose is to draw a drone toward node \( n \) for data collection. Hence, both \( b_{\text{origin}} \) and \( b_{\text{dest}} \) are set to \( b(n) \), so that the order is fully characterized by a single target location instead of an origin-destination pair. The timestamp \( t^{o}_s = t \) records the epoch at which the order is generated, while \( t^{o}_d = t + T_m\Delta T \) marks its expiration: if the order remains unmatched when the next monitoring epoch begins, it is discarded and \( n \) is re-evaluated against the refreshed monitoring scores. Finally, \( \kappa_o = \text{monitor} \) tags the order's type, distinguishing it from real food orders (\( \kappa_o = \text{food} \)) within the unified order set \( \mathcal{O}_t \) so that the shared bipartite matching program and value network in subsection~\ref{subsec_Heterogenous_BM} can apply type-specific feasibility constraints while still evaluating both types on a common scale. This construction converts monitoring, originally a node-level routing objective evaluated at every hop, into a destination-level task defined once per matching epoch.

These virtual orders are then pooled with the real food-delivery orders \( \mathcal{O}_{L,t} \) into a unified heterogeneous order set \( \mathcal{O}_t = \mathcal{O}_{L,t} \cup \mathcal{O}_{Z,t} \), over which a single bipartite match between drones and orders is performed. Consequently, the joint matching-and-routing decision collapses into one unified action---\emph{which order (real or virtual) to serve next}---while the microscopic flight paths are delegated to a deterministic shortest-path router (e.g., OSRM~\citep{ProjectOSRM}), which solves a deadline-feasible insertion-based routing problem over each drone's assigned stops. Formally, this reduces the per-agent decision cardinality from \( O(|\mathcal{N}(n_{m,t})|^H) \) under raw multi-hop routing over an \(H\)-step horizon to \( O(|\mathcal{O}_t|) \) under order-level matching, without sacrificing monitoring coverage: drones continue to accrue monitoring bonuses $B_{m,z,t}$ over the road segments incident to every node they reach en route, so delivery trips opportunistically contribute to network sensing as well.

With the Monitoring-task orderization formulated, the components of the drone agent are then defined as:

1) {\em State}: The state \( s_{m,t} \) of drone \( m \) at time \( t \) is:
\begin{equation}
    s_{m,t} = (I_{m,t}, p_{m,o_l, t}, z_{m,t})
\end{equation}
where $I_{m,t}$ is the information of drone $m$ as previously stated in Equation~\eqref{eq:drone_info}, \( p_{m, o_l, t} \)  represents the information sets of the observed incoming food orders within the matching distance of agent \( m \), \( z_{m,t} \) encodes the observed candidate virtual monitoring orders, including their node locations and current aggregate monitoring scores, which enables the agent to evaluate the long-term value of monitoring tasks against delivery tasks. 

2) {\em Action}: For an available drone \( m \) at time \( t \), after observing the unified order set, the platform assigns an action deciding which candidate order (real food order or virtual monitoring tasks) to serve:
\begin{equation}
\label{eq:action}
\begin{aligned}
    a_{m,t} &= p_{m,o,t} \\
            &= (b_{\text{origin}}, b_{\text{dest}}, t_s^o, t_d^o, \kappa_o),
\end{aligned}
\end{equation}
where \( o \in \mathcal{O}_t = \mathcal{O}_{L,t} \cup \mathcal{O}_{Z,t} \) is the potential order assigned to drone \( m \), and \( \kappa_o \in \{\text{food}, \text{monitor}\} \) indicates the order type. For a food order \( o \in \mathcal{O}_{L,t} \), \( b_{\text{origin}} \in \mathcal{B} \) is the originating launchpad zone, \( b_{\text{dest}} \in \mathcal{B} \) is the destination kiosk zone, \( t_s^o \in \mathbb{R} \) is the time the order was placed, and \( t_d^o \in \mathbb{R} \) is the expected delivery deadline. For a virtual monitoring order \( o \in \mathcal{O}_{Z,t} \), \( b_{\text{origin}} \) and \( b_{\text{dest}} \) are both set as the zone of the selected top-\(k\) monitoring node. \( t_s^o \) is the generated time of the virtual monitoring order node and \( t_d^o \) is the epoch expiration time, after which unserved virtual orders are discarded and re-generated with refreshed monitoring scores. At each time step \( t \), the food order set \( \mathcal{O}_{L,t} \) is updated with newly arrived orders following a stochastic arrival process and assumed to be delivered by couriers directly to the nearest designated launchpad, while the virtual order set \( \mathcal{O}_{Z,t} \) is refreshed every \( T_m \Delta T \) minutes as described above. When the drone is unavailable for matching due to communication failure, the order metadata \( p_{m,t} \) is set to a dummy value \( (b', b', 0, 0, \varnothing) \), where \( b' \) is a placeholder zone index, indicating that no valid order is matched and the drone will remain idle or continue its current service tasks.

3) {\em Reward Function}: \blue{Once matched, the selected order node is inserted into the drone's onboard route via the deadline-feasible insertion VRP solved by the OSRM router~\citep{ProjectOSRM}. Importantly, the detour induced by a newly assigned task is evaluated explicitly rather than statistically approximated. After insertion, OSRM recomputes the updated route, from which we obtain the revised arrival time of each onboard order. The additional delay imposed on every existing order is therefore calculated as the difference between its updated arrival time and its previously planned arrival time, and any resulting deadline violation is directly reflected in the penalty term \(P_{m,t}\). Based on the updated route, the reward \(r_{m,t}\) for drone \(m\) at time \(t\) is then calculated to instantiate the per-agent decomposition of the centralized objective in Equation~\eqref{eq:centralized_objective}:}
\begin{equation}~\label{eq:reward}
     r_{m,t} = B_{m,o_l,t} + \alpha B_{m,z,t} - C_{m,t} - P_{m,t}
\end{equation}
The reward formulation effectively captures the competing objectives, where the incentives of earning monitoring bonuses $B_{m,z,t}$ and maximizing new order bundling $B_{m,o_l,t}$ through routing detours could potentially conflict with the objective of ensuring timely order deliveries enforced by $P_{m,t}$ and drone utilization cost $C_{m,t}$. Note that when a virtual monitoring order enters this formulation, it carries no delivery bonus of its own: when $\kappa_o = \mathrm{monitor}$, the term $B_{m,o_l,t}$ evaluates to zero. Its value is instead realized indirectly and over time. Once the target node is inserted into the drone's route by the deadline-feasible router, the drone passes through a different sequence of road network nodes than it otherwise would, and the road segments incident to these newly reached nodes are credited through $B_{m,z,\tau}$ at every subsequent epoch $\tau \ge t$ along the modified path, while the induced detour is charged immediately through $C_{m,t}$ and $P_{m,t}$. Assigning a monitoring task therefore alters the reward stream not by adding a term to Equation~(\ref{eq:reward}), but by reshaping the trajectory that generates it. 

4) {\em Learning Objective}: By allowing drone agents to share the same allocation policy \( \pi_z \), drone agents aim to find the optimal shared policy \( \pi_z^* \) to maximize their total accumulative rewards via:
\begin{equation}
    \pi_z^{*} = \argmax_{\pi_z} \mathbb{E}_{s_{m}, g_{m} \sim P, a_{m} \sim \pi_z} \left[ \sum_{m \in \mathcal{M}} \sum_{\tau=0}^{T} \gamma^\tau \cdot r_{m,t + \tau} \mid s_{m,t}, g_{m,t}, a_{m,t} \right]
\end{equation}
where \( \pi_z \) is the shared assignment policy mapping states \( s_{m,t} \) and coordination graph \( g_{m,t}\) to actions \( a_{m,t} \). Specifically, \( g_{m,t} = \{\mathcal{V}_{m,t}, \mathcal{E}_{m,t}\} \) denotes the localized bipartite matching coordination graph of agent $m$ at time $t$, where \( \mathcal{V}_{m,t} \) represents the set of agent nodes contained in the local graph, and \( \mathcal{E}_{m,t} \) denotes the corresponding set of edges. Edges in this graph are only established between the ego agent (i.e., agent \( m \)) and other agents if those agents lie within a predefined bipartite matching radius, whose details will be introduced in Section~\ref{sec_GraphMARL}. Accordingly, we construct a global adjacency matrix for the platform whose rows and columns are both indexed by the platform's drone agents. \( \mathbb{E}_{s_{m,t}\sim P, a_{m,t} \sim \pi_z} \) denotes the expectation under policy function \( \pi_z \) and the unknown global state transition function \( P \), \( \gamma \) is the discount factor, and \( r_{m,t + \tau} \) is the reward for drone \( m \) at time \( t + \tau \). This objective can also be expressed in terms of the state-action value function \( Q_{\pi_z}(s_{m,t}, a_{m,t}) \), which represents the expected cumulative reward for taking action \( a_{m,t} \) in state \( s_{m,t} \) and following policy \( \pi_z \) thereafter:
\begin{equation} \label{eq:MAMDP value function}
    \pi_z^{*} = \argmax_{\pi_z} \sum_{m \in \mathcal{M}} Q_{\pi_z}(s_{m,t}, g_{m,t}, a_{m,t})
\end{equation}

\begin{remark}\label{remark:policy}
The behavior policy of each drone agent is not implemented via independent action sampling from \( p(a_{m,t} | s_{m,t}, g_{m.t} )\); instead, \( \pi_z \) is exactly instantiated as a two-stage process: (i) \emph{decentralized graph interdependent value evaluation}, where each agent evaluates its candidate agent-order pairs via the shared Q-function \( Q_{\pi_z}(s_{m,t}, g_{m,t}, a_{m,t}) \), and (ii) \emph{centralized constrained execution}, where a dynamic heterogeneous bipartite matching converts these decentralized value assessments into a conflict-free, globally feasible joint action. When the matching weights are set to \( w(m,o) = Q_{\pi_z}(s_{m,t},  g_{m,t}, a_{m,t}) \) or related variants, the optimal matching solution recovers the constrained joint-greedy maximization of Equation~\eqref{eq:MAMDP value function}, subject to the feasibility constraints detailed below.
\end{remark}

\subsubsection{Dynamic Heterogeneous Bipartite Match Process Augmented with MAMDP Evaluations}\label{subsec_Heterogenous_BM}

We propose a collaborative task assignment framework where the final executed action of each drone agent is exactly the output of the optimal heterogeneous bipartite matching solution, whose weights are derived from the agent's individual MAMDP value evaluation. In other words, the bipartite matching process serves as the centralized execution module of the shared policy \( \pi_z \) in Remark~\ref{remark:policy}. Formally, the dynamic heterogeneous bipartite matching process at time \( t \) is defined as the following constrained ILP problem:
\begin{subequations} \label{eq:ILP}
\begin{align}
& \underset{y_t}{\text{max}}
\sum_{m \in \mathcal{M}} \Big( \sum_{o \in \mathcal{O}_{L,t}} w(m,o) \cdot y_{m,o,t} + \sum_{o \in \mathcal{O}_{Z,t}} w(m,o) \cdot y_{m,o,t} \Big), \label{drone_match_obj}\\
\text{s.t.} \quad
& \sum_{m:m\in \mathcal{M}} y_{m,o,t} \leq 1, \quad \forall o\in \mathcal{O}_t,\label{drone_match_order} \\
& \sum_{o:o\in \mathcal{O}_t} y_{m,o,t} \leq 1, \quad \forall m\in \mathcal{M}, \label{drone_match}\\
& \sum_{o:o\in \mathcal{O}_{L,t}} y_{m,o,t} \leq v_{m,t}, \quad \forall m\in \mathcal{M}, \label{drone_capacitymatch}\\
& \sum_{m:m\in \mathcal{M}} y_{m,o,t}\cdot d_{m,o} \leq R_{match}, \quad \forall o\in \mathcal{O}_{L,t}, \label{drone_match_distance}\\
& y_{m,o,t} \in \{0,1\}, \quad \forall m\in \mathcal{M},o\in \mathcal{O}_{t},
\end{align}
\end{subequations}
where \( y_t=\{y_{m,o,t}\}_{m\in \mathcal{M},o\in \mathcal{O}_t} \) denotes the set of decision variables for the drone fleet over the unified heterogeneous order set \( \mathcal{O}_t = \mathcal{O}_{L,t} \cup \mathcal{O}_{Z,t} \). The objective \eqref{drone_match_obj} maximizes the platform's total expected value gains across both order types. Crucially, solving delivery and monitoring assignments within a single ILP, rather than in two decoupled silos, allows the matching to internalize their competition explicitly: a drone is assigned a virtual monitoring order only when its long-term value \( Q_{\pi_z} \) outweighs the opportunity cost of serving a food order, a trade-off that rule-based or decoupled dispatching schemes cannot capture. Constraint \eqref{drone_match_order} ensures that each order (real or virtual) is matched with at most one drone, which guarantee the satisfaction of real-world implementation requirements. Constraint \eqref{drone_match} guarantees that each drone is matched with at most one delivery or monitoring order at each decision time; note that order bundling is still achieved across successive matching epochs, as a drone with vacant capacity can be incrementally assigned new orders at later epochs and each epoch could be quickly solved, with constraint~\eqref{drone_capacitymatch} guaranteeing that the drone is never overloaded with real delivery orders. Constraint~\eqref{drone_match_distance} guarantees that a drone and a real food delivery order can be matched only if the distance between them \( d_{m,o} \) is within a maximum matching distance \( R_{match} \), which enhances the system's real-time performance.

To endow the shared policy \( \pi_z \) with the standard reinforcement learning exploration-exploitation tradeoff, we design the matching weight \( w(m,o) \) (i.e., the individual value evaluation of drone \( m \) for order \( o \)) using an \( \epsilon \)-greedy strategy, where the randomness is applied at the individual agent level to align with the decentralized evaluation paradigm:
\begin{equation} \label{eq:Matching Value Function}
\begin{cases}
& w(m,o)=  Q_{\pi_z}(s_{m,t}, g_{m,t}, a_{m,t}), \quad  \text{with probability } 1 -\epsilon  \\
& w(m,o)=\overline{Q}, \quad \text{with probability } \epsilon
\end{cases}
\end{equation}
where \( Q_{\pi_z}(s_{m,t},g_{m,t}, a_{m,t}) \) is the previously defined value function in Eq.~\eqref{eq:MAMDP value function} that measures the expected cumulative reward gains, and \( \overline{Q} \) is a large positive number. Since a sufficiently large \( \overline{Q} \) dominates the ILP objective, the perturbed agent-order pair is forced into the optimal matching whenever feasible, thereby realizing exploration at the individual-agent level while preserving the global feasibility guaranteed by the ILP constraints \eqref{drone_match_order}--\eqref{drone_match_distance}. The trade-off between exploitation and exploration is controlled by the parameter \( \epsilon \in (0,1) \) that specifies the exploration rate.

\section{Graph-MAQL Framework for Large-Scale Drone Fleet Coordination under Heterogeneous Task Assignments}~\label{sec_GraphMARL}

In this section, we present our solution approach for the above decentralized, graph-interdependent MAMDP with heterogeneous bipartite matching, aiming to learn an optimal policy $\pi_z$. Building upon the graph-based MARL framework in~\citep{11125816}, we develop a Graph Multi-Agent Q-Learning (Graph-MAQL) framework to coordinate large-scale drone fleets in the presence of heterogeneous order tasks. Specifically, Subsection~\ref{subsec_BM_Graph} describes the construction and encoding of the localized, interdependent bipartite matching graph. Subsection~\ref{subsec_Graph_MAQL} then presents the Graph-MAQL value-evaluation architecture used for large-scale drone fleet coordination and heterogeneous task assignment, while the detailed graph-attention operations and training procedure are provided in the appendices.

\subsection{Construction and Encoding of Localized Interdependent Bipartite Match Graph}~\label{subsec_BM_Graph}

At time \( t \), when coordinating the drone agent fleet, the interdependence among agents primarily emerges from the order assignment process: agents within the pickup range of the same food-delivery orders may compete for identical orders, while agents in proximity to the same road network nodes may contend for the same virtual monitoring orders. Motivated by this observation, for each agent \( m \), we define a localized bipartite matching coordination graph as:
\begin{equation}
    g_{m,t} = \{\mathcal{V}_{m,t}, \mathcal{E}_{m,t}\}
\end{equation}
where \( \mathcal{V}_{m,t} \) denotes the set of agent nodes contained in the localized graph, and \( \mathcal{E}_{m,t} \) denotes the corresponding set of edges. In this graph, an edge is established between the ego agent (i.e., agent \( m \) itself) and another agent only if the latter falls within a predefined proximity threshold, referred to as the bipartite matching radius. At the platform level, as shown in the Graph-based MARL framework of Figure~\ref{fig_frameworkoverview}, we construct a global adjacency matrix whose rows and columns are both indexed by the drone agents on the order-pooling platform. Specifically, the entry \( (i, j) \) of this matrix is set to 1 if agent \( i \) and agent \( j \) share the same order observation or the distance between them is within the bipartite matching radius, and to 0 otherwise. Consequently, the adjacency structure of the localized graph \( g_{m,t} \) can be directly retrieved from the \( m \)-th row (or, by symmetry, the \( m \)-th column) of the platform-level matrix. Accordingly, we denote by \( \mathcal{N}(m) \) the neighbor set of agent \( m \) within the matching radius, which includes drone \( m \) itself.


\subsection{Graph-MAQL Value Evaluation with Localized Graph Encoding}
\label{subsec_Graph_MAQL}

Building upon the localized bipartite matching graph introduced above, we
develop Graph-MAQL to explicitly capture the matching-induced interdependence
among drone agents. For each ego agent $m$, a multi-layer graph attention
encoder aggregates information from its localized matching neighborhood,
allowing the learned representation to reflect both direct competition for
candidate tasks and higher-order dependencies propagated through neighboring
agents. The detailed graph-attention operations, including the input
projection, multi-head query-key-value attention, neighborhood aggregation,
and residual updates, are provided in Appendix~\ref{app_gat_details}.

Let $h_{m,t}^{(0)}$ denote the initial embedding of ego agent $m$ and
$g'_{m,t}$ denote its final coordination-aware graph representation produced
by the $L$-layer graph attention encoder. We concatenate the two representations
as
\begin{equation}
\label{eq:GATDDQN_fused_representation}
    \bar{h}_{m,t}
    =
    h_{m,t}^{(0)}
    \,\Vert\,
    g'_{m,t}.
\end{equation}
This fused representation preserves the ego agent's own operational state
while incorporating the matching-induced dependencies encoded by its localized
bipartite graph.

For each feasible candidate task $o$, the fused representation is further
combined with the fixed-dimensional action feature $\phi(a_{m,o,t})$ and fed
into a shared nonlinear value network:
\begin{equation}
\label{eq:DDQN_backbone}
    Q_{\theta}
    \left(
        s_{m,t},
        g_{m,t},
        a_{m,o,t}
    \right)
    =
    f_{\mathrm{Q}}
    \left(
        \bar{h}_{m,t}
        \,\Vert\,
        \phi(a_{m,o,t});
        \theta_{\mathrm{Q}}
    \right),
\end{equation}
where $\theta$ includes the parameters of both the graph encoder and the
downstream Q-network. Since the heterogeneous attributes of food-delivery and
virtual monitoring orders are already embedded in the unified state-action
representation, the same value function is shared across all drone agents and
both task types.

The resulting Graph-MAQL therefore serves as a shared evaluator of all feasible
agent-task pairs. At each matching epoch, its Q-value replaces the corresponding
edge weight in Equation~\eqref{eq:Matching Value Function} of the heterogeneous
bipartite matching problem:
\begin{equation}
\label{eq:GATDDQN Matching Value Function}
\begin{cases}
w(m,o)
=
Q_{\theta}
\left(
s_{m,t},
g_{m,t},
a_{m,o,t}
\right),
& \text{with probability } 1-\epsilon,\\
w(m,o)=\overline{Q},
& \text{with probability } \epsilon.
\end{cases}
\end{equation}
The matching program then converts these coordination-aware pairwise values
into a globally feasible joint assignment. In this way, Graph-MAQL combines
decentralized value estimation with centralized constraint-feasible execution,
while retaining scalability through parameter sharing and localized graph
aggregation.

\paragraph{Graph-MAQL training.} The complete training procedure, including the temporal-difference updates, target-network stabilization, exploration strategy, and training pseudocode, is provided in Appendix~\ref{app:graph_maql_training}.

\section{Case Study and Discussions}\label{sec_Simulation_Results}
\subsection{Simulation Setup}
{(a) \em Dataset and study area:} To validate the proposed framework, we construct a realistic simulation environment for the Kowloon region of Hong Kong using food-delivery demand, road-congestion data, and road-network information. Historical meal-delivery data from Foodpanda and Deliveroo~\citep{FoodpandaHK, DeliverooHK} provide trip-specific information, including order time, pickup location, drop-off location, food item, and restaurant information. Road speeds are constructed by integrating official Hong Kong traffic-speed data~\citep{HK_OfficialData} with OpenStreetMap (OSM)~\citep{ProjectOSRM} and HereMap~\citep{HEREMAP_API}. Details of the speed-map fusion and graph-based smoothing procedure, together with visualizations of the food-delivery demand and road-speed data, are provided in Appendix~\ref{app_simulation_details}. The resulting Kowloon road network contains 2,388 intersection nodes and 4,114 road segments over approximately 47~km$^2$.

{(b) \em Simulation Setup:} To thoroughly investigate the synergy between on-demand food delivery and monitoring objectives, as well as the scalability of our algorithmic framework, we configured a simulation involving 200 drones, focusing on the peak period from 17:30 to 19:30. We assume that kiosks and launchpads are abundantly distributed throughout the study area, so that every order can be associated with a nearby launchpad (for courier handoff) and a nearby kiosk (for customer pickup). For each order assigned to drone service, we assume that couriers initially deliver the order to the nearest launchpad as quickly as possible, after which it is transported by a drone to the nearest kiosk. For orders not serviced by drones within 5 minutes, couriers are dispatched to complete the delivery. Each drone is assigned a carrying capacity \(c_m = 3\) orders. Our simulation spans a time horizon of 120 minutes, with decision intervals \(\Delta t = 1\) minute and road half-life period $T_h$ for each road segment is set to 10 minutes to reflect dynamic traffic conditions. At the start of each simulation episode, drones are randomly distributed across the network. The expected average delivery time window $t_d^o$ is set as 40 minutes, and the base price for accepting an order \(R_o\) is 30 HKD, with a distance-based bonus scaling factor \(\beta_d = 5\) HKD/km. Operational costs per drone, \(C_m\), are set at 0.1 HKD/min. The discount factor \(\gamma\) is configured to 0.99. From the customer’s perspective, virtual penalty coefficients for extra detours, \(\lambda_c\) and \(\lambda_p\), are set at 0.1 HKD/min and 3 HKD/min, respectively. For bipartite matching, the maximum allowable matching distance \(R_{\text{match}}\) for real delivery order is set to 2 km. For routing optimization, instead of allowing drones to fly along direct routes, we constrain drones to fly above the road network only. This adjustment accounts for the dense high-rise buildings in Hong Kong and the associated monitoring requirements. We set the speed of couriers as 25 km/h and utilized the Open Source Routing Machine (OSRM) model to solve the VRP after the order assignments, which provides real-time route guidance based on the actual urban road network. This approach facilitates accurate estimations of onboard times and pickup schedules for the orders.

{(c) \em Algorithm configuration:} \blue{The monitoring score in Equation~(\ref{eq:monitoring_score}) balances congestion and freshness with $\alpha_c = 0.5$, and the top-$k$ orderization is executed every $T_m \Delta T = 5$~minutes. In our experiments, we consider \(k \in \{0,25,50,75,100\}\), spanning settings from no virtual monitoring orders to relatively aggressive monitoring-task injection. For the localized bipartite matching coordination graph, the bipartite matching radius is set to $2$~km, and the fixed-budget graph sampling retains the ego agent together with at most $20$ sampled neighbors, padding the neighborhood with masked dummy nodes whenever fewer agents are available. Each layer of our graph attention encoder stacks includes $K = 3$ attention heads, with a latent dimension of $F' = 128$ and $d_k = d_v = 64$ per head, while the downstream value network is a three-layer perceptron with $256$, $128$, and $64$ hidden units and $\mathrm{ReLU}$ activations. Training runs for $500$ episodes using the Adam optimizer with a learning rate of $\alpha_\theta = 1 \times 10^{-3}$, drawing mini-batches of $1024$ transitions from a replay memory of capacity $2 \times 10^{4}$ agent-level transitions. The target network is updated by Polyak averaging with $\rho_\theta = 5 \times 10^{-3}$. The exploration rate is initialized at $\epsilon = 1.0$ and decayed exponentially at the beginning of each episode at a rate of $\beta = 0.98$ until reaching the floor value $\epsilon_T = 0.001$. All experiments are conducted on a workstation equipped with an Intel Core i9-14900KF CPU and an NVIDIA RTX 4090 GPU. Under this configuration, a full decision epoch for $200$ drones takes $0.8$~s on average, of which $0.58$~s is spent on value evaluation over all feasible agent-task pairs, $0.01$~s on solving the heterogeneous bipartite matching program, and $0.21$~s on route re-computation by the router, well within the $\Delta t = 1$~minute decision interval.}

\subsection{Synergy between Drone Delivery Pooling and Road Network Monitoring}\label{sec_synergy}
To demonstrate the synergy between the dual objectives of drone food delivery pooling and road network monitoring, we conduct two groups of experiments under a large-scale configuration featuring 200 drones operating at 54 km/h, with approximately 4300 orders placed over the two-hour peak period through drone delivery pooling mode. The purpose of this subsection is to establish that the synergy is a property of the problem itself rather than an artifact of any particular learning architecture; accordingly, both groups adopt an Independent-MAQL agent under the bipartite matching pipeline, deliberately setting aside the coordination mechanism of Graph-MAQL, whose comparison against alternative baselines is deferred to Section~\ref{sec_baseline_comparison}. The first group isolates the intrinsic synergy by varying the monitoring priority weight \(\alpha\) while disabling the monitoring-task orderization module, so that monitoring influences drone behavior solely through the reward signal. Building on the operating point identified therein, the second group activates orderization and examines how the number of virtual monitoring orders generated per epoch further extends the attainable synergy. For each configuration, agents are trained for 500 episodes and evaluated over 20 independent test episodes. Values are reported as \(x \pm y\), where \(x\) is the mean over test episodes and \(y\) the standard deviation.

\blue{The evaluation metrics reported in Tables~\ref{tab:alpha_sweep} and~\ref{tab:orderization} are defined as follows. \emph{Orders Picked Up} denotes the total number of customer delivery requests successfully accepted and served by the drone fleet over the evaluation horizon, capturing the system's order-serving capacity. \emph{Deadline Violation Rate} is the percentage of picked-up orders delivered beyond their prescribed service deadlines and therefore reflects delivery timeliness and service reliability. \emph{Link Coverage Rate} is defined as the fraction of road links monitored at least once during the evaluation period, measuring the spatial breadth of road-network sensing. \emph{Bottleneck Roads Monitor Frequency} captures how frequently the drone fleet revisits persistently congested road segments. Specifically, for each road link, we compute its average congestion score over the episode based on the observed traffic states and classify the link as a bottleneck road if this episode-level average score is no smaller than the predefined threshold of \(0.6\). The metric is then calculated as the total number of drone visits to all identified bottleneck links divided by the number of such links. \emph{Delivery Score} represents the delivery-side contribution to the system objective and is computed as \(\sum_{t \in \mathcal{T}} \sum_{m \in \mathcal{M}} \left(B_{m,o_l,t} - P_{m,t}\right)\), where the first term rewards successful order fulfillment and the second penalizes deadline violations. \emph{Monitoring Score} represents the cumulative monitoring value generated by the drone fleet through visits to informative road-network nodes and is computed as \(\sum_{t \in \mathcal{T}} \sum_{m \in \mathcal{M}} B_{m,z,t}\), with \(B_{m,z,t}\) determined by the monitoring score defined in Equation~(\ref{eq:monitoring_score}). \emph{Average Occupied Seats}, reported in Table~\ref{tab:alpha_sweep}, denotes the average number of delivery orders simultaneously carried by an active drone and serves as an indicator of the degree of delivery pooling. Finally, \emph{Overall Score}, reported in Table~\ref{tab:orderization}, combines the delivery and monitoring components and corresponds directly to the centralized objective defined in Equation~\eqref{eq:centralized_objective}, thereby providing an aggregate measure of integrated system performance.}

\begin{table}[!htbp]
\centering
\footnotesize
\setlength{\tabcolsep}{4pt}
\caption{Performance under varying monitoring priority weight \(\alpha\) without monitoring-task orderization.}
\label{tab:alpha_sweep}
\begin{tabular}{cccccccc}
\toprule
\(\alpha\) & \makecell{Orders\\Picked Up} & \makecell{Deadline Violation\\Rate (\%)} & \makecell{Link Coverage\\Rate (\%)} & \makecell{Bottleneck Roads\\Monitor Frequency} & \makecell{Delivery\\Score (HKD)} & \makecell{Monitoring\\Score} & \makecell{Average\\Occupied Seats} \\
\midrule
0  & 4188 \(\pm\) 12.2 & 5.0 \(\pm\) 0.90 & 63.9 \(\pm\) 1.07 & 3.90 \(\pm\) 0.18 & 79639 \(\pm\) 657  & 5555 \(\pm\) 145   & 2.25 \(\pm\) 0.03 \\
3  & 4191 \(\pm\) 12.3 & 4.3 \(\pm\) 0.80 & 69.7 \(\pm\) 0.69 & 4.29 \(\pm\) 0.14 & 80129 \(\pm\) 557  & 5926 \(\pm\) 97.3  & 2.31 \(\pm\) 0.03 \\
6  & 4189 \(\pm\) 14.4 & 6.4 \(\pm\) 1.16 & 71.1 \(\pm\) 1.04 & 4.59 \(\pm\) 0.15 & 79057 \(\pm\) 774  & 6269 \(\pm\) 132.3 & 2.33 \(\pm\) 0.04 \\
8  & 4161 \(\pm\) 16.2 & 6.6 \(\pm\) 1.15 & 72.3 \(\pm\) 0.70 & 4.80 \(\pm\) 0.13 & 78129 \(\pm\) 652  & 6550 \(\pm\) 74.3  & 2.40 \(\pm\) 0.03 \\
10 & 4138 \(\pm\) 13.3 & 9.0 \(\pm\) 0.84 & 72.7 \(\pm\) 0.61 & 4.90 \(\pm\) 0.15 & 75747 \(\pm\) 1027 & 6558 \(\pm\) 104.3 & 2.45 \(\pm\) 0.03 \\
\bottomrule
\end{tabular}
\end{table}

Table~\ref{tab:alpha_sweep} reports the first group, in which monitoring enters the system only through the reward signal. Two observations stand out. 

First, at a moderate weight of \(\alpha = 3\), the two objectives improve \emph{jointly} rather than trading off against each other. Relative to the delivery-only baseline \(\alpha = 0\), the Link Coverage Rate rises from 63.9\% to 69.7\% and the Bottleneck Roads Monitor Frequency from 3.90 to 4.29, while delivery performance does not deteriorate at all: orders picked up remain statistically indistinguishable (4191 versus 4188), the Delivery Score improves from 79639 to 80129, and the Deadline Violation Rate in fact \emph{declines} from 5.0\% to 4.3\%. The Average Occupied Seats, which rises from 2.25 to 2.31, identifies the mechanism behind this counterintuitive gain. The monitoring reward pulls drones away from over-competed launchpad clusters, and the resulting spatial dispersion exposes each drone to a broader set of pending orders, so that more of them can be consolidated onto a single trip. Denser pooling reduces the effective travel burden per order, which more than offsets the detour incurred for monitoring. The synergy is therefore mediated by pooling: monitoring incentives do not merely coexist with delivery but actively reshape the spatial distribution of the fleet in a manner that favors order bundling. Notably, this effect emerges under an independent-learning agent without any explicit inter-agent coordination, confirming that it originates from the structure of the integrated problem rather than from a particular algorithmic design.

Second, reward shaping alone saturates rapidly. As \(\alpha\) increases beyond 3, monitoring gains become progressively more expensive: the Delivery Score falls monotonically to 75747 at \(\alpha = 10\), a 4.9\% loss relative to the baseline, while the Deadline Violation Rate nearly doubles to 9.0\%. More revealingly, the monitoring objectives themselves plateau, the Monitoring Score moves only from 6550 to 6558 between \(\alpha = 8\) and \(\alpha = 10\), and the Link Coverage Rate stalls near 72\% despite a further 3\% sacrifice in delivery performance. The reason is structural: monitoring credit accrues only at the nodes a drone actually reaches, and the reward signal offers no mechanism to actively steer a drone toward a distant high-value node before it arrives there. Monitoring thus remains a passive by-product of delivery-driven trajectories, and network coverage is ultimately bounded by the spatial footprint of the delivery demand itself. Overcoming this ceiling requires making monitoring an assignable task in its own right, which is precisely the role of monitoring-task orderization.

\begin{table}[!htbp]
\centering
\footnotesize
\setlength{\tabcolsep}{5pt}
\caption{Performance under varying numbers \(k\) of virtual monitoring orders generated every five minutes, with monitoring-task orderization enabled and \(\alpha\) fixed at 3. The row \(k=0\) coincides with the \(\alpha=3\) configuration of Table~\ref{tab:alpha_sweep}.}
\label{tab:orderization}
\begin{tabular}{cccccccc}
\toprule
\makecell{$k$} & \makecell{Orders\\Picked Up} & \makecell{Deadline\\Violation\\Rate (\%)} & \makecell{Link\\Coverage\\Rate (\%)} & \makecell{Bottleneck\\Roads Monitor\\Frequency} & \makecell{Delivery\\Score (HKD)} & \makecell{Monitoring\\Score} & \makecell{Overall\\Score (HKD)} \\
\midrule
0   & 4191 \(\pm\) 12.3 & 4.3 \(\pm\) 0.80 & 69.7 \(\pm\) 0.69 & 4.29 \(\pm\) 0.14 & 80129 \(\pm\) 557  & 5926 \(\pm\) 97  & 97887 \(\pm\) 710 \\
25  & 4203 \(\pm\) 14.4 & 5.6 \(\pm\) 1.19 & 74.2 \(\pm\) 0.75 & 5.13 \(\pm\) 0.20 & 80105 \(\pm\) 1016 & 6936 \(\pm\) 111 & 100894 \(\pm\) 902 \\
50  & 4197 \(\pm\) 15.4 & 5.9 \(\pm\) 0.73 & 75.9 \(\pm\) 0.63 & 5.68 \(\pm\) 0.19 & 79834 \(\pm\) 888  & 7248 \(\pm\) 78  & 101556 \(\pm\) 881 \\
75  & 4194 \(\pm\) 15.1 & 6.3 \(\pm\) 1.14 & 75.2 \(\pm\) 0.53 & 5.94 \(\pm\) 0.18 & 80099 \(\pm\) 904  & 7361 \(\pm\) 70  & 102159 \(\pm\) 893 \\
100 & 4168 \(\pm\) 15.8 & 6.8 \(\pm\) 0.68 & 75.2 \(\pm\) 1.00 & 6.09 \(\pm\) 0.16 & 79362 \(\pm\) 642  & 7413 \(\pm\) 78  & 101576 \(\pm\) 708 \\
\bottomrule
\end{tabular}
\end{table}

\blue{Fixing \(\alpha = 3\) as the operating point identified above, the second group activates monitoring-task orderization and varies the number \(k\) of virtual monitoring orders instantiated every five minutes. We consider \(k\in\{0,25,50,75,100\}\), spanning from no virtual monitoring demand to a relatively aggressive monitoring configuration. Relative to the realized food-delivery demand in our experiments, \(k=50\) corresponds to approximately 25\% of the number of actual delivery orders generated over the same interval, whereas \(k=100\) raises this ratio to roughly 50\%. Table~\ref{tab:orderization} shows that orderization fundamentally reshapes the trade-off frontier.}

Monitoring performance improves markedly and continues improving well past the ceiling observed in Table~\ref{tab:alpha_sweep}. As \(k\) grows from 0 to 100, the Monitoring Score rises by 25.1\% from 5926 to 7413 and the Bottleneck Roads Monitor Frequency by 42.0\% from 4.29 to 6.09, while the Link Coverage Rate reaches 75.9\% at \(k=50\), a level unattainable through reward shaping alone even at \(\alpha=10\). These gains come at negligible delivery cost: the Delivery Score varies within 1\% across all settings (80129 to 79362), orders picked up remain essentially unchanged, and the Deadline Violation Rate increases by only 2.5 percentage points.

The contrast between the two mechanisms is stark. Under reward shaping alone, lifting the Monitoring Score by 18.1\% (5555 to 6558) required pushing \(\alpha\) to 10 and forfeiting 4.9\% of the Delivery Score. With orderization, a larger 25.1\% monitoring improvement costs less than 1\% of delivery performance, and the Bottleneck Roads Monitor Frequency reaches 6.09 versus the 4.90 attainable through reward weighting. The explanation is structural: by instantiating high-score nodes as explicitly assignable tasks, the platform can \emph{direct} spare fleet capacity toward under-monitored regions through the matching program, instead of waiting for delivery trajectories to incidentally overfly them. Monitoring is thereby elevated from a passive by-product of delivery routing into an actively schedulable objective, while the shared matching program continues to ensure that a drone accepts a monitoring task only when its value exceeds the opportunity cost of the delivery order it displaces.

A closer look at the Overall Score reveals an interior optimum at \(k = 75\) (102159), beyond which the aggregate declines to 101576 at \(k = 100\) despite the Monitoring Score continuing to rise. The composition of monitoring effort explains this reversal. As more virtual orders are generated per epoch, an increasing share of them concentrates on persistently congested corridors that dominate the score ranking, so the fleet revisits critical bottlenecks ever more frequently---Bottleneck Roads Monitor Frequency climbs steadily from 5.13 to 6.09---while breadth across peripheral links ceases to improve, with the Link Coverage Rate flattening at 75.2\% for \(k \geq 75\). Past this point the marginal monitoring gain no longer compensates the accumulating delivery cost, as reflected in the Deadline Violation Rate rising to 6.8\% and the Delivery Score dropping to its lowest value in the table. This indicates the existence of a well-defined operating point at which the two objectives are best balanced, and confirms that \(k\) offers the operator a single interpretable dial for navigating between uniform network coverage and intensive surveillance of critical corridors.

\subsection{Comparison of Proposed Graph-MAQL Framework against Baselines}
\label{sec_baseline_comparison}

We now turn to the coordination mechanism itself and show that Graph-MAQL improves coordination along two dimensions simultaneously: \emph{among agents}, by resolving contention for a shared and rapidly depleting order pool, and \emph{between the two objectives}, by pricing delivery and monitoring tasks against each other more accurately. These are not separate mechanisms but two facets of a single estimation problem, since the matching weight \(w(m,o)\) must encode both judgments at once.

We fix $\alpha = 3$ and deliberately adopt $k = 100$ rather than the aggregate optimum $k = 75$ identified in Section~\ref{sec_synergy}. The rationale is that the decline observed beyond $k = 75$ originates precisely from delivery capacity being over-committed to monitoring under an independent-learning dispatcher: it is the regime in which uncoordinated evaluation degrades most severely, and therefore the regime in which the value of coordination is most clearly exposed. Holding this operating point fixed, we vary only the coordination mechanism, so that performance differences are attributable solely to how inter-agent interdependence is handled. Three methods are compared: (i) \emph{Greedy}, a myopic rule-based dispatcher that assigns each drone to the feasible task with the highest immediate reward~\citep{dong2026real}, isolating the contribution of long-horizon value estimation; (ii) \emph{Independent-MAQL}, the independent-learning agent adopted throughout Section~\ref{sec_synergy}, in which each drone evaluates candidate agent-task pairs without observing its neighbors, following the spirits of~\citep{enders2023hybrid,maljkovic2025learning}; and (iii) the proposed \emph{Graph-MAQL}. All three share the same heterogeneous bipartite matching pipeline, the same monitoring-task orderization module, and the same reward structure, differing only in how the matching weights are produced. The two learning-based methods are trained under identical budgets and random seeds.

\blue{We evaluate under two demand levels. The first, denoted \(1.0\times\), corresponds to the baseline volume of approximately 4,300 orders on which both learning-based policies are trained. Both Independent-MAQL and Graph-MAQL exhibit stable convergence after approximately 300 training episodes, well before the 500-episode training budget used in our experiments. The second, denoted \(1.25\times\), raises the volume to approximately 5,300 orders while holding the fleet size at 200 drones, thereby tightening capacity. Crucially, no policy is retrained for this second level: both Independent-MAQL and Graph-MAQL are evaluated \emph{zero-shot}, executing the policies learned at \(1.0\times\) in an environment whose demand density they have never encountered. This protocol serves two purposes. It stresses the interdependence that Graph-MAQL is designed to capture, since contention for overlapping order pools intensifies as capacity tightens; and it probes whether the relational inductive bias of the graph attention encoder confers robustness to shifts in fleet-level operating conditions, a property of practical consequence, as demand on real platforms fluctuates far faster than policies can be retrained. Greedy, being rule-based, requires no training and is therefore unaffected by this distinction; it is reported at both levels for reference.}

\begin{table}[!htbp]
\centering
\footnotesize
\setlength{\tabcolsep}{4pt}
\caption{Comparison of Graph-MAQL against baselines. Both learning-based policies are trained at the \(1.0\times\) demand level; panel (b) reports zero-shot evaluation of those same policies under approximately 5,300 orders, with no retraining. The Independent-MAQL row in panel (a) coincides with the \(k=100\) configuration of Table~\ref{tab:orderization}. Abbreviations: Viol.\ Rate = deadline violation rate; Link Cov.\ = link coverage rate; Bttlnk.\ Freq.\ = bottleneck roads monitoring frequency.}
\label{tab:baseline_comparison}

\vspace{0.5em}
{\small (a) \(1.0\times\) demand (\(\approx\) 4,300 orders, training level)}
\vspace{0.3em}

\begin{tabular}{lccccccc}
\toprule
Method & \makecell{Orders\\Picked Up} & \makecell{Viol.\\Rate (\%)} & \makecell{Link Cov.\\(\%)} & \makecell{Bttlnk.\\Freq.} & \makecell{Delivery\\Score} & \makecell{Monitor.\\Score} & \makecell{Overall\\Score} \\
\midrule
Greedy              & 3702 \(\pm\) 15.8 & 19.0 \(\pm\) 1.26 & 76.8 \(\pm\) 0.79 & 6.27 \(\pm\) 0.23 & 63238 \(\pm\) 753 & 7617 \(\pm\) 81 & 86059 \(\pm\) 906 \\
Independent-MAQL    & 4168 \(\pm\) 15.8 & 6.8 \(\pm\) 0.68  & 75.2 \(\pm\) 1.00 & 6.09 \(\pm\) 0.16 & 79362 \(\pm\) 642 & 7413 \(\pm\) 78 & 101576 \(\pm\) 708 \\
\textbf{Graph-MAQL} & \textbf{4199 \(\pm\) 6.8} & \textbf{3.9 \(\pm\) 0.39} & 74.0 \(\pm\) 0.75 & 5.84 \(\pm\) 0.16 & \textbf{82332 \(\pm\) 225} & 7231 \(\pm\) 103 & \textbf{104000 \(\pm\) 378} \\
\bottomrule
\end{tabular}

\vspace{1.2em}
{\small (b) \(1.25\times\) demand (\(\approx\) 5,300 orders, zero-shot)}
\vspace{0.3em}

\begin{tabular}{lccccccc}
\toprule
Method & \makecell{Orders\\Picked Up} & \makecell{Viol.\\Rate (\%)} & \makecell{Link Cov.\\(\%)} & \makecell{Bttlnk.\\Freq.} & \makecell{Delivery\\Score (HKD)} & \makecell{Monitor.\\Score} & \makecell{Overall\\Score (HKD)} \\
\midrule
Greedy              & 3815 \(\pm\) 14.8 & 21.7 \(\pm\) 1.66 & 76.3 \(\pm\) 0.49 & 6.10 \(\pm\) 0.13 & 64611 \(\pm\) 546 & 7601 \(\pm\) 74 & 87387 \(\pm\) 522 \\
Independent-MAQL    & 4641 \(\pm\) 29.8 & 5.1 \(\pm\) 0.58  & 74.6 \(\pm\) 0.72 & 6.51 \(\pm\) 0.18 & 88755 \(\pm\) 857 & 7621 \(\pm\) 77 & 111596 \(\pm\) 822 \\
\textbf{Graph-MAQL} & \textbf{4713 \(\pm\) 22.2} & \textbf{2.9 \(\pm\) 0.53} & 73.8 \(\pm\) 0.80 & 6.27 \(\pm\) 0.18 & \textbf{91940 \(\pm\) 513} & 7542 \(\pm\) 73 & \textbf{114544 \(\pm\) 545} \\
\bottomrule
\end{tabular}
\end{table}

Table~\ref{tab:baseline_comparison} reports the comparison, from which three observations follow.

\textbf{Long-horizon value estimation is indispensable on the delivery side.} The Greedy dispatcher performs worst by a wide margin on every delivery metric, picking up 3702 orders against 4199 for Graph-MAQL at the \(1.0\times\) level and incurring a Deadline Violation Rate of 19.0\% versus 3.9\%. The failure mode is instructive: by assigning each drone to whichever task offers the largest immediate reward, Greedy accepts orders whose insertion into an existing route is locally attractive but ultimately jeopardizes the deadlines of on-board orders, and the resulting cascade of violations depresses the Delivery Score to 63238. Since Greedy operates on the same matching pipeline and the same orderization module as the learning-based methods, this gap confirms that the performance reported throughout this section stems from learned value estimation rather than from the surrounding architecture alone.

\textbf{Coordination among agents improves delivery performance, and the advantage survives a shift in demand density.} At the \(1.0\times\) level on which both policies are trained, Graph-MAQL raises the Delivery Score by 3.7\% over Independent-MAQL (82332 versus 79362) while nearly halving the Deadline Violation Rate from 6.8\% to 3.9\%. Executed zero-shot at \(1.25\times\), the advantage is preserved rather than eroded: the Delivery Score improves by 3.6\% (91940 versus 88755) and the violation rate again falls by almost half, from 5.1\% to 2.9\%, with 72 additional orders completed against 31 at the trained level. Two properties of the architecture account for this robustness. First, the improvement itself arises from resolving contention---under an independent policy a drone evaluates a candidate order without knowing whether nearby drones are contending for the same order or are already carrying detour-constrained loads, and the graph attention encoder recovers the value lost to this miscalibration by conditioning each estimate on the states of neighbors within the matching radius. That the violation rate is roughly halved at both demand levels indicates the gain comes specifically from avoiding assignments whose detours would cascade into deadline failures. Second, because attention is computed over a neighborhood of bounded size and the same parameters are shared across all agents, a denser demand regime alters the \emph{content} of each agent's neighborhood without altering the structure of the computation, so the learned coordination policy transfers without retraining. The markedly smaller standard deviations of Graph-MAQL, for instance \(\pm\)225 against \(\pm\)642 in Delivery Score at \(1.0\times\), further indicate that coordinated evaluation yields not only higher but also more stable fleet-level outcomes.

\textbf{Coordination between the two objectives settles at a better point on the frontier.} Graph-MAQL attains slightly lower Link Coverage and Bottleneck Roads Monitor Frequency than Independent-MAQL, while Greedy attains the highest monitoring scores of all three despite its collapse in delivery performance. This ordering follows directly from how each method prices the two task types within the shared matching program. The value of a monitoring task is realized immediately upon traversal and depends only on the road network state, whereas the value of a delivery order is contingent on the future: on whether the induced detour will jeopardize on-board deadlines, and on whether competing drones will absorb the remaining orders. A myopic dispatcher therefore systematically overprices monitoring relative to delivery, and Greedy's superior monitoring metrics are a symptom of this mispricing rather than a strength---capacity that delivery needed more has been diverted to sensing, at the cost of a 19.0\% violation rate. Independent-MAQL corrects part of this bias through learned value estimation, but its estimates remain miscalibrated where contention is severe. Graph-MAQL, pricing both task types most accurately of the three, settles at an allocation that trades a marginal amount of monitoring---1.2 and 0.8 percentage points of Link Coverage at the two demand levels---for delivery gains an order of magnitude larger. The Overall Score, which weights the two objectives by the same \(\alpha = 3\) used during training, confirms this as the better operating point at both levels: 104000 against 101576 at \(1.0\times\), and 114544 against 111596 under zero-shot evaluation at \(1.25\times\).

Taken together, these results show that the two dimensions of coordination reinforce rather than compete with each other. Sharper value estimates simultaneously resolve which drone should serve a contested order and whether that order is worth more than the monitoring task it displaces, because both judgments are carried by the same matching weight. The resulting allocation dominates both the myopic and the independent-learning baselines on the aggregate objective, and does so without retraining when demand density rises by a quarter---a form of robustness that matters in deployment, where order volumes fluctuate far faster than dispatch policies can be relearned.

\section{Conclusions}\label{sec_conclusions}

This work studies the real-time dispatch of a shared drone fleet for jointly serving on-demand food delivery pooling and urban road network monitoring. We propose \emph{monitoring-task orderization}, which converts road-network monitoring demand into virtual monitoring orders and pools them with real food orders into a unified heterogeneous task set. This abstraction transforms the coupled matching-and-routing problem into an order-level decision process while allowing delivery and monitoring objectives to be jointly evaluated. Building on this formulation, we develop Graph-MAQL, a decentralized graph-interdependent Multi-Agent Markov Decision Process in which localized bipartite match coordination graphs capture inter-agent contention and a shared value network evaluates heterogeneous delivery and monitoring tasks on a common scale. A dynamic heterogeneous bipartite matching program then converts these decentralized value estimates into globally feasible joint assignments.

Extensive experiments on a realistic Hong Kong Kowloon simulation yield three findings. First, delivery and monitoring are mutually reinforcing rather than merely compatible: at a moderate monitoring weight, network coverage improves substantially while delivery performance does not deteriorate at all, an effect mediated by order pooling, as the monitoring reward disperses drones away from over-competed launchpad clusters and thereby exposes each drone to a broader set of poolable orders. Second, reward shaping alone saturates rapidly, since an edge is credited only after it has been traversed and monitoring therefore remains a passive by-product of delivery-driven trajectories; orderization overcomes this ceiling by elevating monitoring into an actively schedulable objective, achieving a 25.1\% gain in monitoring score at less than 1\% delivery cost. Third, Graph-MAQL improves coordination along two dimensions simultaneously—among agents, by resolving contention for a shared and rapidly depleting order pool, and between the two objectives, by pricing them against each other more accurately—and these gains transfer zero-shot to a demand density a quarter higher than that seen during training, a form of robustness that matters in deployment where order volumes fluctuate far faster than dispatch policies can be relearned.

Several directions merit further investigation. First, the periodic top-\(k\) selection introduces an approximation relative to the centralized formulation: nodes ranked just below the threshold receive no explicit incentive until scores are refreshed, and a learned or adaptive generation policy that conditions \(k\) and the node selection on real-time fleet availability could recover part of the coverage that remains unattained even at high monitoring priority. Second, the current formulation assumes homogeneous drones with identical payload capacity and endurance; extending the heterogeneous matching program to typed agent classes, and strengthening the coordination graph through relational graph convolution networks~\citep{schlichtkrull2018modeling}, would accommodate mixed fleets with differentiated delivery and sensing capabilities. Third, deployability would benefit from incorporating urban airspace regulations through geofencing constraints, dynamic weather conditions that modulate flight parameters, and an energy-aware extension in which battery state-of-charge enters the state representation and charging actions enter the action space, so that the framework balances operational efficiency against energy consumption and the sustainability of drone-based urban services.

\appendix

\section{Detailed Graph Attention Encoding in Graph-MAQL}
\label{app_gat_details}

This appendix provides the detailed graph-attention operations underlying the
coordination-aware representation used in Graph-MAQL. These standard attention
operations are omitted from the main text for conciseness.

Building upon the localized graph structure introduced above, we integrate a multi-layer graph attention network (GAT) with the localized bipartite matching graph to capture the matching-induced interdependence among agents. The resulting architecture is illustrated in the Graph-MAQL framework of Figure~\ref{fig_frameworkoverview}. Let $\mathcal{N}_{t}(i)$ denote the set of neighboring nodes of node $i$ in this graph. We first project the state feature 
of each node into a common latent space using a shared linear transformation, similar to the feature projection employed in GraphSAGE~\citep{hamilton2017inductive}:
\begin{equation}
\label{eq:GAT_input_projection}
    h_{i,t}^{(0)} = W_{\mathrm{in}} s_{i,t},
\end{equation}
where $s_{i,t} \in \mathbb{R}^{F}$ is the original state feature of node $i$, 
$h_{i,t}^{(0)} \in \mathbb{R}^{F'}$ is its initial latent representation, and 
$W_{\mathrm{in}} \in \mathbb{R}^{F' \times F}$ is a learnable projection matrix 
shared across all nodes.

To capture higher-order dependencies in the localized matching graph, we stack 
$L$ graph attention layers. At the $\ell$-th GAT layer, where 
$\ell \in \{1,\ldots,L\}$, the query, key, and value vectors $q_{i,t}^{(\ell,k)}$, $k_{j,t}^{(\ell,k)}$, and $v_{j,t}^{(\ell,k)}$ associated with the 
$k$-th attention head are computed as
\begin{equation}
\label{eq:GAT_QKV}
\begin{aligned}
    q_{i,t}^{(\ell,k)}
        &= W_{Q}^{(\ell,k)} h_{i,t}^{(\ell-1)}, \\
    k_{j,t}^{(\ell,k)}
        &= W_{K}^{(\ell,k)} h_{j,t}^{(\ell-1)}, \\
    v_{j,t}^{(\ell,k)}
        &= W_{V}^{(\ell,k)} h_{j,t}^{(\ell-1)},
\end{aligned}
\end{equation}
where $W_{Q}^{(\ell,k)}, W_{K}^{(\ell,k)} \in \mathbb{R}^{d_k \times F'}$ and $W_{V}^{(\ell,k)} \in \mathbb{R}^{d_v \times F'}$ are learnable projection matrices. 

Following the scaled dot-product attention mechanism
in~\citep{vaswani2017attention}, the unnormalized compatibility score between
node $i$ and its neighbor $j$ at graph attention layer $\ell$ and attention
head $k$ is computed as
\begin{equation}
\label{eq:GAT_attention_score}
    e_{ij,t}^{(\ell,k)}
    =
    \frac{
        \left(q_{i,t}^{(\ell,k)}\right)^{\mathrm{T}}
        k_{j,t}^{(\ell,k)}
    }{
        \sqrt{d_k}
    }.
\end{equation}
The superscript $\mathrm{T}$ denotes matrix transposition, and $d_k$ is the dimension of the query and key vectors. The
scaling factor $\sqrt{d_k}$ prevents the magnitude of the dot product from
growing excessively with the feature dimension, thereby improving numerical
stability during training.

The compatibility scores are subsequently normalized over the neighborhood of
node $i$ using the softmax function:
\begin{equation}
\label{eq:GAT_attention_coefficient}
    \alpha_{ij,t}^{(\ell,k)}
    =
    \frac{
        \exp\left(e_{ij,t}^{(\ell,k)}\right)
    }{
        \displaystyle
        \sum_{u \in \mathcal{N}_{t}(i)}
        \exp\left(e_{iu,t}^{(\ell,k)}\right)
    },
    \qquad
    j \in \mathcal{N}_{t}(i),
\end{equation}
where $\exp(\cdot)$ denotes the exponential function and
$\mathcal{N}_{t}(i)$ denotes the set of neighboring nodes whose information
is accessible to node $i$ at time $t$. The resulting coefficient
$\alpha_{ij,t}^{(\ell,k)}$ quantifies the relative importance of neighbor $j$
to node $i$ under attention head $k$, satisfying
\begin{equation}
    \alpha_{ij,t}^{(\ell,k)} \geq 0,
    \qquad
    \sum_{j\in\mathcal{N}_{t}(i)}
    \alpha_{ij,t}^{(\ell,k)} = 1.
\end{equation}
For each attention head, the neighborhood information is aggregated as
\begin{equation}
\label{eq:GAT_single_head_aggregation}
    h_{i,t}^{(\ell,k)}
    =
    \sum_{j \in \mathcal{N}_{t}(i)}
    \alpha_{ij,t}^{(\ell,k)}
    v_{j,t}^{(\ell,k)}.
\end{equation}
The outputs of the $K$ attention heads are concatenated and projected back to 
the latent feature space:
\begin{equation}
\label{eq:GAT_multi_head_aggregation}
    h_{i,t}^{(\ell)}
    =
    \sigma\left(
        W_{O}^{(\ell)}
        \left(
            \big\Vert_{k=1}^{K}
            h_{i,t}^{(\ell,k)}
        \right)
        +
        W_{R}^{(\ell)}h_{i,t}^{(\ell-1)}
    \right),
\end{equation}
Here, $\Vert$ denotes vector concatenation,
$W_{O}^{(\ell)} \in \mathbb{R}^{F_{\ell} \times Kd_v}$ is the multi-head
output projection matrix, and
$W_{R}^{(\ell)} \in
\mathbb{R}^{F_{\ell} \times F_{\ell-1}}$
projects the preceding-layer representation onto the current latent feature
space. Moreover, $\sigma(\cdot)$ denotes a nonlinear activation function.
The residual term preserves the node-specific information from the preceding
layer while the multi-head aggregation term incorporates information from its
graph neighbors. By recursively applying Equation~\eqref{eq:GAT_multi_head_aggregation} over $L$
graph attention layers, information is progressively propagated across the
localized matching graph. Consequently, the final-layer embedding
$h_{i,t}^{(L)}$ captures both the node's own state information and the
multi-hop matching dependencies encoded by its graph neighborhood.

For the localized matching graph centered on ego agent $m$, we extract the
final-layer embedding of its corresponding agent node as the
matching-aware graph representation:
\begin{equation}
\label{eq:GAT_graph_representation}
    g'_{m,t} = h_{m,t}^{(L)}.
\end{equation}
Through the graph attention layers, each 
agent can exploit both direct matching relations and higher-order dependencies 
in the localized bipartite matching graph, thereby facilitating more coordinated 
delivery-and-monitoring behaviors among agents with heterogeneous tasks.

\section{Training Graph-MAQL for Large-Scale Drone Fleets with Heterogeneous Task Assignments}
\label{app:graph_maql_training}
During training, the transition generated by agent $m$ at time $t$ is
represented as
\begin{equation}
\label{eq:Graph-MAQL_transition}
    \tau_{m,t}
    =
    \left(
        s_{m,t},
        g_{m,t},
        a_{m,t},
        r_{m,t},
        s_{m,t+1},
        g_{m,t+1},
        a_{m,t+1},
        d_{m,t+1}
    \right),
\end{equation}
where $s_{m,t}$ and $g_{m,t}$ denote the local state and localized matching
graph observed by agent $m$, respectively; $a_{m,t}$ is the action assigned
to agent $m$ by the heterogeneous bipartite matching process; and $r_{m,t}$
is the resulting immediate reward. The binary variable
$d_{m,t+1}\in\{0,1\}$ is the terminal indicator, where $d_{m,t+1}=1$
indicates that the episode terminates upon reaching decision epoch $t+1$. For a nonterminal transition, $a_{m,t+1}$ denotes the action assigned to
agent $m$ at the subsequent decision epoch by the behavior policy, including
both the agent-level evaluation mechanism and the centralized constrained
matching process in our heterogeneous bipartite matching process. Retaining $a_{m,t+1}$ is essential as Graph-MAQL
adopts a SARSA-style temporal-difference target, which evaluates the
subsequent action actually generated by the behavior policy rather than an
independently selected greedy action. For a terminal transition, no valid
subsequent action exists, and $a_{m,t+1}$ is assigned the dummy value.
The dummy action does not affect the temporal-difference target because the
bootstrapped term is masked by $1-d_{m,t+1}$.

To stabilize temporal-difference learning, Graph-MAQL maintains an online
network with parameters $\theta$ and a target network with parameters
$\theta^-$. The one-step target is defined as
\begin{equation}
\label{eq:GraphMAQL_SARSA_target}
    y_{m,t}
    =
    r_{m,t}
    +
    \gamma
    \left(1-d_{m,t+1}\right)
    Q_{\theta^-}
    \left(
        s_{m,t+1},
        g_{m,t+1},
        a_{m,t+1}
    \right),
\end{equation}
where $\gamma\in[0,1)$ is the discount factor. For a terminal transition,
$d_{m,t+1}=1$, and the target reduces to $y_{m,t}=r_{m,t}$.

The online Graph-MAQL network is trained by minimizing the mean-squared
temporal-difference error:
\begin{equation}
\label{eq:GraphMAQL_loss}
    \mathcal{L}(\theta)
    =
    \mathbb{E}_{\tau_{m,t}\sim\mathcal{D}}
    \left[
        \left(
            y_{m,t}
            -
            Q_{\theta}
            \left(
                s_{m,t},
                g_{m,t},
                a_{m,t}
            \right)
        \right)^2
    \right],
\end{equation}
where $\mathcal{D}$ denotes the replay buffer containing the collected
agent-level transitions. Specifically, let
\begin{equation}
\label{eq:GraphMAQL_gradient}
    \mathbf{g}
    =
    \nabla_{\theta}\mathcal{L}(\theta)
\end{equation}
denote the gradient of the loss with respect to the training network
parameters. The training-network parameters are then updated using
the clipped gradient:
\begin{equation}
\label{eq:GraphMAQL_online_update}
    \theta
    \leftarrow
    \theta
    -
    \alpha_{\theta}
    \mathbf{g},
\end{equation}
where $\alpha_{\theta}$ denotes the learning rate. The target network parameters
are subsequently updated through Polyak averaging:
\begin{equation}
\label{eq:GraphMAQL_target_update}
    \theta^-
    \leftarrow
    \rho_{\theta}\theta
    +
    \left(1-\rho_{\theta}\right)\theta^-,
\end{equation}
where $\rho_{\theta}\in(0,1)$ is the target network update rate and is
typically assigned a small value.

The complete training procedure of the proposed Graph-MAQL framework is
summarized in Algorithm~\ref{al: Graph-MAQL Framework}. Steps 1 and 2
initialize the simulation environment, including the router
model~\citep{ProjectOSRM}, the drone fleet configuration and the underlying
road network, together with the replay memory, the training and target
network parameters $\theta$ and $\theta^-$, and the associated training
hyper-parameters. The framework then enters the training phase, in which each
episode begins by generating the episodic order requirements in step 4. To
achieve a balance between exploitation and exploration during the training
process, the exploration rate of the bipartite matching process is decayed at
the beginning of every episode in step 5 via:
\begin{equation} \label{eq:epsilon_decay}
\epsilon = \max(\epsilon \cdot \beta, \epsilon_T)
\end{equation}
where $\beta$ is the exponential decay rate and $\epsilon_T$ is the final
exploration rate. This gradual reduction in the exploration rate is designed to
shift the focus from exploration to exploitation as the learning advances.

Within each episode, steps 7 to 13 are executed at every decision epoch with a
time interval of $\Delta t$. In step 7, the central platform refreshes the
system-level information, i.e., the pending food-delivery orders, the location
and on-board order status of each drone, and the monitoring score distribution
over the road network. Based on the updated monitoring scores, step 8 performs
the top-$k$ orderization that converts the monitoring demand of the road
network into a set of virtual monitoring orders, so that the delivery and
monitoring tasks can be handled within our heterogeneous bipartite match assignment process. In
step 9, the platform constructs the localized bipartite matching graph for each
agent, evaluates every feasible agent-order pair with the shared Graph-MAQL
value network according to Equation~\eqref{eq:GATDDQN Matching Value Function},
and solves the resulting heterogeneous bipartite matching problem in
Equation~\eqref{eq:ILP} to obtain the assignment decisions. Steps 10 to 12 constitute the learning loop. In step 10, the drone agents
execute their assigned actions in the platform, and the transition tuple
$\tau_{m,t}$ defined in Equation~\eqref{eq:Graph-MAQL_transition} of every agent,
which stores the local state, the localized matching graph, the assigned action,
the immediate reward and the subsequent state-graph-action triplet required by
the SARSA-style target, is appended to the replay memory. In step 11, a
mini-batch $\mathcal{D}$ is sampled from the replay memory and the training
network parameters $\theta$ are updated with the clipped gradient in
Equation~\eqref{eq:GraphMAQL_online_update}, whereas step 12 softly updates the
target network parameters $\theta^-$ through the Polyak averaging in
Equation~\eqref{eq:GraphMAQL_target_update}. Finally, in step 13, the central
platform recomputes and updates the flight route of each drone according to the
performed action, which in turn determines the system evolution at the next
decision epoch.

\begin{algorithm}
\caption{Graph-MAQL Framework} \label{al: Graph-MAQL Framework}
\begin{algorithmic}[1] 
\STATE Simulator Initialization: Router Model~\citep{ProjectOSRM}, Drones Fleet Configurations, Road Network. 
\STATE Graph-MAQL Initialization: Experience Memory, Training Net Parameter $\theta$, Target Net Parameter $\theta^-$, and Training Hyper-parameters $\alpha_\theta$, $\rho_\theta$, Discount Rate $\gamma$, Exploration Rate $\epsilon$, $\epsilon_T$ with Exponential Decay Rate $\beta$.
\FOR{$e = 1$ to Episodes}
    \STATE Initialize: Episode Order Requirements
    \STATE Perform exponential decay according to Equation~(\ref{eq:epsilon_decay})
    \FOR{$t = 0$ to $t_{\text{terminal}}$ by $\Delta t$}
        \STATE Central platform updates order information, each drone’s location, on-board order status, and road network monitoring score distributions.
        \STATE Central platform conduct top-k road network monitoring task orderization.
        \STATE Central platform assigns heterogeneous order tasks to drone agents according to formulation in Equations (\ref{eq:GATDDQN Matching Value Function}) and (\ref{eq:ILP}).
        \STATE Drones observe their orders, perform their assigned actions in the platform and add every agent’s new experience tuple $\tau_{m,t}$ as in Equation~\eqref{eq:Graph-MAQL_transition} into the memory.
        \STATE Sample experience tuples as mini-batch $D$ and use Equation~(\ref{eq:GraphMAQL_online_update}) to update $\theta$.
        \STATE Update target network parameters $\theta^-$ using Equation~(\ref{eq:GraphMAQL_target_update}).
        \STATE Based on the performed action, central platform calculates and update the new route.
    \ENDFOR
\ENDFOR
\end{algorithmic}
\end{algorithm}

\section{Road-Speed Data Preprocessing and Simulation Data Visualization}
\label{app_simulation_details}
The road-congestion environment is constructed using a fused speed map that integrates official Hong Kong traffic-speed data~\citep{HK_OfficialData}, OpenStreetMap (OSM)~\citep{ProjectOSRM}, and HereMap~\citep{HEREMAP_API}. Official speed observations are first assigned to OSM road edges through nearest-neighbor matching. For road segments without official observations, HereMap speeds are used to supplement the network. To obtain a spatially complete and topologically consistent speed field, we further apply a graph-based iterative smoothing procedure based on the node-edge incidence structure of the road network. Road segments associated with official measurements are treated as fixed anchor edges, while the speeds of non-fixed edges are iteratively updated using the average speeds of their topological neighbors. The procedure terminates after 10 iterations or when no further changes in edge speeds are observed. This process produces a complete road-speed field that remains consistent with both available traffic measurements and the underlying network topology.

\begin{figure}[!htbp]
    \centering
    \subfigure[Food Delivery Dataset]{
        \includegraphics[width=0.9\textwidth, height=8.5cm]{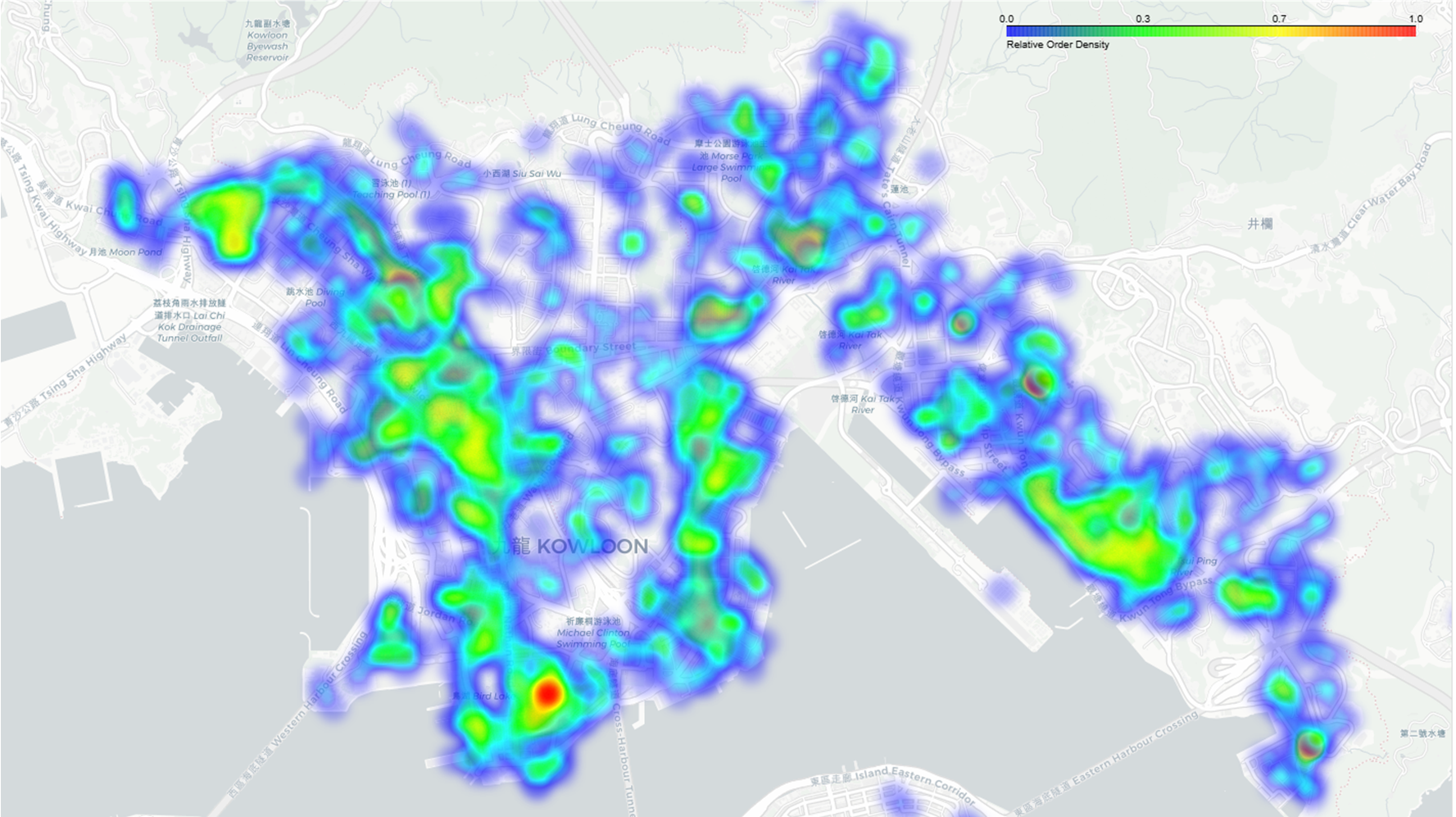}
        \label{fig:Food Delivery Dataset}
    }
    \subfigure[Road Network Speed Map]{
        \includegraphics[width=0.9\textwidth, height=8.5cm]{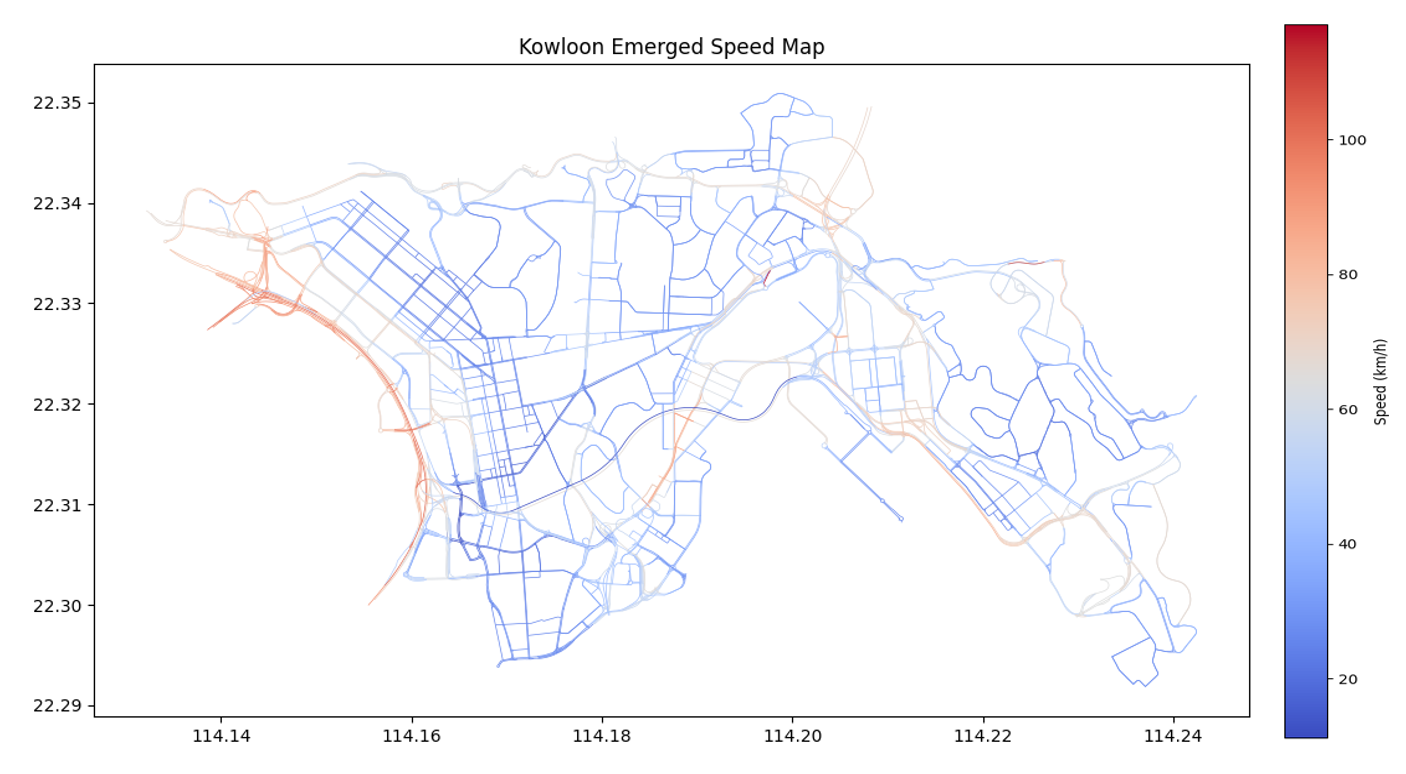}
        \label{fig:Speedmap}
    }
    \caption{Visualization of the Kowloon food-delivery dataset and road-network speed map.}
    \label{fig:appendix_data_visualization}
\end{figure}

Figure~\ref{fig:appendix_data_visualization} visualizes the spatial distributions of the food-delivery demand and the resulting road-network speed map used in the simulation.

\bibliographystyle{plainnat}
\bibliography{sample-base}

\end{document}